# Non-Inferiority and Equivalence Tests in Sequential, Multiple Assignment, Randomized Trials (SMARTs)

Palash Ghosh[1], Inbal Nahum-Shani[2], Bonnie Spring[3], and Bibhas Chakraborty[1,4,5]

[1]Centre for Quantitative Medicine, Duke-NUS Medical School, National University of Singapore, Singapore

[2]Institute for Social Research, University of Michigan

[3]Center for Behavior and Health, Northwestern University Feinberg School of Medicine

[4]Department of Statistics and Applied Probability, National University of Singapore

[5]Department of Biostatistics and Bioinformatics, Duke University

emails:

palash.ghosh@duke-nus.edu.sg,

inbal@umich.edu

bspring@northwestern.edu

bibhas.chakraborty@duke-nus.edu.sg

Correspondence concerning to this article should be addressed to Palash Ghosh, Centre for Quantitative Medicine, Duke-NUS Medical School, Level 6, Academia, 20 College Road, Singapore 169856. Email: palash.ghosh@duke-nus.edu.sg

**Abstract**

Adaptive interventions (AIs) are increasingly becoming popular in behavioral sciences. An AI is a sequence of decision rules that specify for whom and under what conditions different intervention options should be offered, in order to address the changing needs of individuals as they progress over time. The *sequential, multiple assignment, randomized trial* (SMART) is a novel trial design that was developed to aid in empirically constructing effective AIs. The sequential randomizations in a SMART often yield multiple AIs that are embedded in the trial by design. Many SMARTs are motivated by scientific questions pertaining to the comparison of such embedded AIs. Existing data analytic methods and sample size planning resources for SMARTs are suitable only for superiority testing, namely for testing whether one embedded AI yields better primary outcomes on average than another. This calls for non-inferiority/equivalence testing methods, since AIs are often motivated by the need to deliver support/care in a less costly or less burdensome manner, while still yielding benefits that are equivalent or non-inferior to those produced by a more costly/burdensome standard of care. Here, we develop data analytic methods and sample size formulas for SMART studies aiming to test the non-inferiority or equivalence of one AI over another. Sample size and power considerations are discussed with supporting simulations, and online sample size planning resources are provided. A simulated data analysis shows how to test a non-inferiority or equivalence hypothesis with SMART data. For illustration, we use an example from a SMART study aiming to develop an AI for promoting weight loss among overweight/obese adults.

**Introduction**

High heterogeneity in response to behavioral interventions motivates the development of adaptive interventions (Collins, Murphy, & Bierman, 2004). In adaptive interventions (AIs), different intervention options (e.g., different types, intensities or modalities of delivery of treatment) are offered based on ongoing information about the individual's changing conditions. An AI is not an experimental design; it is an intervention design (i.e., the approach and specifics of an intervention program; see Nahum-Shani et al., 2017) that seeks to address the unique and changing needs of individuals as they progress over time through an intervention program. An AI (also known as a dynamic treatment regime in the statistical literature; see, e.g., Murphy, 2003; Chakraborty & Moodie, 2013; Kosorok & Moodie, 2016) is a protocol that guides for whom and under what conditions different intervention options should be offered, typically operationalized using a sequence of decision rules involving some tailoring variables. AIs have been discussed, investigated, and applied across various domains of psychological sciences, including health psychology (e.g., Nahum-Shani et al., 2015; Thomas & Bond, 2015), clinical psychology (e.g., Connell, Dishion, Yasui, & Kavanagh, 2007; Pelham et al., 2016), educational psychology (e.g., Schaughency & Ervin, 2006; Walkington, 2013), and organizational psychology (Howard & Jacobs, 2016; Eden, 2017).

The development of an AI should be guided by evidence (empirical, theoretical and/or practical). However, in many cases existing evidence is insufficient to inform the development of effective AIs; scientific questions might arise concerning which intervention option to offer, for whom, and under what conditions. The *sequential, multiple assignment, randomized trial* (SMART; Lavori & Dawson, 2000, 2004; Thall, Millikan, & Sung, 2000; Murphy, 2005) is a novel experimental design that was developed to aid in the construction of empirically grounded AIs. In a SMART, some or all participants are randomized more than once, with the sequential randomizations typically resulting in several AIs that are embedded in the SMART by design. Many SMARTs are motivated by scientific questions concerning the comparison of these embedded AIs.

As the importance of developing evidence-based AI is increasingly acknowledged in many areas of behavioral and psychological research, SMART designs are experiencing rapid uptake despite being relatively new. For example, in the area of clinical psychology, SMARTs were employed to develop AIs for promoting weight loss among African American adolescents (Naar-King et al., 2016), to develop dynamic strategies for re-engaging individuals in treatment (McKay et al., 2015), to optimize the use of contingency management in the treatment of substance use (NIH/NIAAA R01AA021446; PI: Petry; Petry et al., 2018), and to develop an adaptive behavioral smoking cessation intervention for people living with HIV (NIH/NIMH R01DA034537; PI: Ledgerwood). In the area of health psychology, SMARTs have been recommended as an experimental tool that can be used to inform the development of adaptive behavior change interventions that capitalize on advances in mobile and wireless technology (Riley et al., 2015). In organizational psychology, Howard and Jacobs (2016) employed a SMART to optimize the sequencing of training methods, Eden (2017) discussed the utility of the SMART as an experimental approach that can inform the adaptation of organizational policies and practices. In their introduction to AIs and SMARTs in the area of educational and school psychology, August, Piehler and Miller (2018) provide multiple examples of SMART applications in this area. Many other examples of psychological research involving or highlighting the scientific utility of the SMART exist in the extant literature (see e.g., Lei et al., 2012; Lu et al., 2016).

Similarly, new methodologies have been developed to enable investigators to analyze data arising from SMART studies (Nahum-Shani et al., 2012a; Chakraborty, Laber, & Zhao, 2013; Chakraborty, Ghosh, Moodie, & Rush, 2016; Laber, Lizotte, Qian, Pelham, & Murphy, 2014; Shortreed, Laber, Stroup, Pineau, & Murphy, 2014; Zhao, Zeng, Laber, & Kosorok, 2015a; Ertefaie, Shortreed, & Chakraborty, 2016) and to plan sample size for SMARTs (Oetting et al., 2011; Kidwell & Wahed, 2013; Laber et al., 2016). Several tutorial manuscripts have provided guidelines specific to psychological and behavioral scientists considering this experimental approach (see Almirall et al., 2014; Nahum-Shani et al., 2012a; Nahum-Shani et al., 2012b; and Lei et al., 2012). These tutorials review various types of SMART designs, offer an accessible introduction to data analytic methods that can be used with data arising from a SMART, and review sample size considerations. However, an important gap exists with respect to these methodologies.

Existing methodologies for analyzing data, and for planning sample size for SMART studies are suitable only for superiority testing, namely for testing whether one embedded AI yields better primary outcome on average than another. However, the development of AIs often requires addressing scientific questions concerning non-inferiority or equivalence. Specifically, many AIs are motivated by the need to deliver support/care in a less costly or less burdensome manner (Lei et al., 2012; Nahum-Shani et al., 2017), while still yielding benefits that are equivalent or non-inferior to those produced by more costly or burdensome interventions that represent the standard of care. While Chuang-Stein, Follman, and Chappell (2014) commented about the potential role of non-inferiority testing in the context of a SMART design, to the best of our knowledge, there has been no follow up work on this topic in terms of developing the actual tests and associated sample size resources. This represents a gap in the science of AIs, given that the motivation to minimize burden and cost compared to established behavioral intervention approaches underlie the development of AIs in various health (Collins, Murphy, & Strecher, 2007; Riley, Rivera, Atienza, Nilsen, Allison, & Mermelstein, 2011), academic (Almirall et al., 2018a), and clinical (Page et al., 2016; Pelham et al., 2016) domains. The goal of this manuscript is to expand the methodological toolbox available to psychological and behavioral researchers considering a SMART by developing data analytic methods and sample size formulas for SMART studies that are motivated by primary scientific questions concerning the non-inferiority or equivalence of one AI over another.

We begin by providing a brief review of AIs and SMART studies followed by other types of SMART designs and commonly used data analysis methods with software tools. After developing the general setup and introducing notation, we discuss data analytic methods and related sample size planning resources for SMARTs that are motivated by non-inferiority testing, and then for SMARTs that are motivated by equivalence testing. Finally, we present simulation studies to validate the proposed methodologies. We also develop an online tool that implements our methodology. Technical details, as well as an illustration of data analysis using simulated data are provided in the Appendix. Throughout, for illustration, we use an example from the SMART Weight Loss Management study (ClinicalTrials.gov Identifier: NCT02997943), which aims to develop an AI that integrates mobile technology into the treatment of overweight/obese adults.

**Adaptive Interventions (AI)**

An AI is a sequence of individualized treatments that use ongoing information about the individual's progress (e.g., early signs of non-response or non-adherence) to decide whether and how the intervention should be modified. An AI consists of four main elements: (a) decision points, namely points in time in which a decision should be made concerning whether and how to intervene; (b) intervention options, which are different types, dosages/intensities or delivery modalities that may be used/offered at any given decision point; (c) tailoring variables, namely baseline and time-varying information about the individual that is useful in deciding the type/dose/modality of the intervention at each decision point; and (d) decision rules, which link the tailoring variables to intervention options, specifying which intervention option to offer, for whom, and under what conditions; there is a decision rule for each decision point (Almirall et al., 2014; Nahum-Shani et al., 2012a). An AI is a multi-stage process, where each stage corresponds to a period of time following a decision point in which the individual experiences the assigned intervention option; the assigned intervention option in at least one of the stages is tailored based ongoing (time-varying) information about the participant, namely information that may change over time as a result of prior intervention stages (e.g., response to prior intervention, level of adherence to initial treatment, or motivational changes during the previous stage) (Nahum-Shani et al., 2017).

As an example, consider an AI for promoting weight loss among overweight/obese adults (Figure 1). At the first stage of this AI, a mobile app is offered to all individuals; this app is designed to support self-monitoring of dietary intake and physical activity. The individual's response status is determined at weeks 2, 4, and 8 based on the amount of weight loss. As long as the individual is losing at least 0.5 lb. on average per week, s/he is classified as a responder; otherwise the individual is classified as a non-responder. As soon as the individual is classified as a non-responder, s/he transitions to the second stage of this AI, wherein additional support is offered in the form of weekly coaching sessions. As long as the individual is responsive, s/he continues with mobile app alone. This intervention is adaptive because it uses ongoing information about the individual's response status to decide whether or not coaching should be added to the mobile weight loss app. The following decision rule specifies the weight loss AI:

At program entry,

    First-stage intervention option=mobile app

Then, at weeks 2, 4, and 8

    If response status = non-responder

       Then, second-stage intervention option=add coaching (and stop assessing response status)

    Else, continue with app alone (and continue assessing response status until week 8).

In this AI, there are decision points at program entry and at weeks 2, 4, and 8. The intervention options include a mobile app (offered to all participants at the first stage), and two subsequent tactics: add coaching and continue with app alone (employed at the second stage depending on the participant's response status). The tailoring variable in this AI is the participant's response status. The decision rule specifies how information about the individual's response status at weeks 2, 4, and 8 should be used to tailor the second-stage intervention options, namely to decide who should receive more support in the form of coaching and who should continue with app alone.

The weight loss AI described above is motivated by considerations pertaining to cost and burden. Coaching is an effective, yet relatively costly and burdensome weight loss treatment component (The Diabetes Prevention Program Research Group, 2003; Appel et al., 2011). The mobile app is less costly and less burdensome. However, empirical evidence suggests that not all individuals benefit sufficiently from using a mobile app to lose weight; about 50% are unlikely to achieve ultimate treatment goals. Importantly, empirical evidence suggests that these individuals can be identified early, based on insufficient weight loss during the first few weeks of mobile app usage (Spring, Pellegrini, & Nahum-Shani, 2014). Hence, instead of offering coaching to all individuals, this AI starts with the less costly/burdensome treatment component (mobile app) and then adds coaching only to those participants who need it most, namely those who show early signs of non-response. As long as the individual is responsive, s/he continues with the less costly/burdensome treatment component. This AI represents a more cost-effective approach than offering coaching sessions to all individuals throughout the intervention program.

More generally, AIs are typically motivated by evidence indicating that the same intervention approach is not beneficial to all individuals in the target population, and that it is possible to identify *early* those individuals who are unlikely to benefit ultimately. AIs are designed to modify the intervention (e.g., increase the intensity of current intervention, switch to another intervention, or augment with another type of intervention) when an individual shows early signs of non-response, in order to prevent ultimate non-response and maximize the number of individuals who benefit from the intervention program. The scientific motivation for AIs and the advantages of this approach over a fixed, one-size-fits-all intervention approach have been extensively discussed in the extant literature (see Collins, Murphy, & Bierman, 2004; Nahum-Shani et al., 2012a; Almirall et al., 2014). Practically, AIs are intended to guide the efforts by therapists, clinicians, teachers and organizations to provide tailored treatments, practices and programs to individuals in order to address their unique and changing needs. The protocolized (i.e., pre-specified) nature of AIs increases their replicability in a real-world implementation, as well as in research aiming to optimize and evaluate their effectiveness.

**The sequential, multiple assignment, randomized trial (SMART)**

The SMART is an experimental trial design that involves multiple stages of randomizations; each stage corresponds to scientific questions concerning the selection and individualization of intervention options at particular decision points in an AI. As an example, consider the SMART weight loss management study for integrating mobile technology in the treatment of obese/overweight adults (R01 DK108678; see Figure 2). At program entry, all individuals are randomized with equal probability (0.5) to one of two first-stage intervention options, either (1) mobile app alone (App) or (2) a mobile app combined with weekly coaching sessions (App+Coaching). Response status is assessed at weeks 2, 4, and 8. As soon as the individual is classified as a non-responder s/he is re-randomized with equal probability (0.5) to one of two second-stage augmentation tactics: either (1) modest augmentation, which consists of adding another technology-based intervention component in the form of supportive text messages, or (2) vigorous augmentation, which consists of adding supportive text messages combined with a more traditional weight loss intervention component (either coaching or meal replacement) that the individual was not offered initially. As long as the individual is classified as responsive, s/he continues with the initial intervention option and is not re-randomized.

The sequential randomizations in the SMART weight loss study are restricted: only non-responders are re-randomized to second stage intervention options, whereas responders are not re-randomized. This restriction is consistent with the scientific questions motivating the weight loss SMART. Specifically, this study was designed to address open scientific questions concerning the selection of front-line treatments and the selection of augmentation tactics for non-responders. For responders, sufficient evidence exists to inform the selection of subsequent tactics. As long as the individual is responsive, the likelihood of achieving ultimate weight loss goals is high and hence continuing with the initial intervention option is both logical (if it's not broken, don't fix it) and resource efficient (Spring et al., 2017; Waring et al., 2014). Because different subsequent intervention options are considered for responders (continue) and non-responders (modest vs. vigorous augmentation), response status is embedded as a tailoring variable in this SMART by design. Such multi-stage restricted randomizations give rise to several AIs that are embedded in the SMART. Table 1 presents the four AIs that are embedded in the SMART weight loss study and the corresponding experimental subgroups that are consistent with each. Notice that a simplified version of one of these AIs was described earlier (in Figure 1). Specifically, embedded in the SMART weight loss study is an AI (AI #1) which, similar to the AI described in Figure 1, recommends initiating treatment with App alone, and then to augment with coaching (and supportive text messaging) as soon as the individual exhibits early signs of non-response, and continue with app alone as long as the individual is responding. Participants in subgroups A and C in Figure 2 are consistent with this embedded AI.

An important scientific question motivating the SMART weight loss study concerns the comparison of AI #1, which recommends coaching only to individuals who show early signs of non-response; and AI #3, which recommends coaching to all individuals in terms of long-term (month 6) weight loss. Specifically, AI #3 recommends to initiate treatment with App and coaching, and then to vigorously augment with supportive text messaging and meal replacement as soon as the individual exhibits early signs of non-response, and to continue with App and coaching as long as the individual is responsive. Participants in subgroups D and F in Figure 2 are consistent with this embedded AI. The rationale for this comparison relates to cost and burden. Because AI #3 recommends coaching throughout, it is likely to be effective, yet relatively costly and burdensome. AI #1, on the other hand, offers coaching only to those individuals who seem to need it most (i.e., early non-responders), hence it is hypothesized to be non-inferior, namely no less effective than AI #3. If AI #1 is equally or more beneficial in terms of ultimate weight loss compared to AI #3, then the former should be selected for real-world implementation because it is less costly and/or burdensome.

**Other Types of SMART Designs**

As noted earlier, given the scientific questions motivating the SMART weight loss study, the sequential randomizations in this SMART are restricted such that only non-responders are re-randomized to second-stage intervention options, whereas responders continue with their assigned initial intervention option. Although this type of design is highly common (Almirall et al., 2018b), other forms of SMARTs exist, depending on the scientific questions motivating the study. These include (a) SMARTs in which the second-stage randomization is not restricted, such that all participants are re-randomized to second-stage intervention options regardless of their early response status (see the remaking recess SMART in Figure 5 of Almirall et al., 2018b); (b) SMARTs in which the second-stage randomization is restricted such that only non-responders to a specific first-stage intervention option (e.g., App alone) are re-randomized to second-stage options, whereas non-responders to the alternative first-stage intervention option (e.g., App + Coaching) as well as all responding participants are offered pre-specified second-stage interventions (see the SMART study for developing an AI for minimally verbal children with autism spectrum disorder in Figure 1 of Almirall et al., 2016); and (c) a SMARTs in which the second-stage randomization is restricted such that different second-stage intervention options are considered for responders vs. non-responders (e.g., responders are re-randomized to two types of maintenance interventions and non-responders are re-randomized to two types of rescue interventions; see the Extending Treatment Effectiveness of Naltrexone SMART study in Figure 1 of Nahum-Shani et al., 2017).

The extant literature on SMARTs offers additional examples and details concerning different types of SMARTs that vary in terms of the re-randomization scheme (see Almirall et al., 2018b; Kidwell, 2016; Lei et al., 2012; Nahum-Shani et al., 2012a) the number of randomization stages (i.e., considering more than two stages of randomization; see Chakraborty et al., 2016), and the level (individual vs. cluster) at which randomizations occur (see NeCamp, Kilbourne, & Almirall, 2017 for a discussion of cluster-randomized SMARTs). For simplicity of illustration, the current manuscript focuses solely on the type of SMART employed in the weight loss study discussed above (Figure 2), namely a SMART that employs two stages of randomization, where all individuals are randomized initially to two first-stage intervention options and only non-responding individuals are re-randomized subsequently to two second-stage intervention options. Given that this type of SMART has become one of the most popular forms of SMARTs, it is often considered the 'prototypical SMART' (see Almirall et al., 2018b; Nahum-Shani & Dziak, 2018; NeCamp et al., 2017).

**Data Analysis Methods and Software Tools**

Data obtained from SMART designs require advanced data analytic methods to analyze them. To compare two adaptive interventions embedded within a SMART, Oetting et al. (2011) and Nahum-Shani et al. (2012a) proposed a z-statistic based approach and an inverse probability weighting approach respectively. In recent years, *Q-learning*, a stage-by-stage recursive estimation procedure (Murphy, 2005; Nahum-Shani et al., 2012b, Chakraborty, Laber & Zhao, 2013) and its variations like interactive Q-learning (Laber, Linn & Stefanski, 2014) have been frequently used to estimate optimal AIs from SMART data. Other noteworthy methods to estimate an optimal AI include classification-based outcome-weighted learning (Zhao et al., 2012, 2015a) and value-search methods based on (augmented) inverse probability weighted estimators (Zhang et al., 2012a, 2012b). Quite a few software packages are available to analyze data according to the various available methods. For example, both the R-package '*qLearn*' and the SAS procedure '*PROC QLEARN*' can be used to estimate optimal AIs using Q-learning and construct associated bootstrap-based confidence intervals, based on two-stage SMART data. More recent R-packages like '*DTRlearn*' and '*DynTxRegime*' implement some of the other methods mentioned above, in addition to Q-Learning, that all aim to estimate the optimal AI.

**Non-Inferiority and Equivalence Testing**

Existing methods for using SMART study data to compare AIs (e.g., Nahum-Shani et al., 2012a) and associated power planning resources (Oetting et al., 2011; Kidwell & Wahed., 2013), are suitable for traditional superiority testing, where the goal is to investigate whether one AI is more efficacious than another. However, in non-inferiority testing, the goal is to establish that a new intervention yields favorable outcomes that, when compared to another active control intervention (i.e., an intervention approach with established evidence of effectiveness), are not below some pre-stated *non-inferiority margin* (Seaman & Serlin, 1998; Piaggio et al., 2006; Wellek, 2010). The non-inferiority margin captures how close the new intervention must be to the other in terms of the expected outcome in order for the new intervention to be considered non-inferior to the other. Practically, the margin is the maximum clinically acceptable difference in favor of the active control intervention that one is willing to accept in return for the secondary benefits (e.g., lower cost, burden and /or side effects) of the new intervention. In the SMART weight loss study, the goal is to test the non-inferiority of AI #1 with respect to AI #3. AI #3 represents an active control as it offers coaching (a treatment component with established evidence of effectiveness) throughout but at the expense of added cost and burden, whereas AI #1 represents a novel approach that initiates treatment with a less costly/burdensome treatment component (a mobile app) and then recommends coaching only to those individuals who need it most (i.e., those showing signs of early non-response). Hence, it is hypothesized that AI #1 is non-inferior in terms of long-term (month 6) weight loss compared to AI #3.

Selecting the non-inferiority margin is the most critical and challenging step in non-inferiority testing. The smaller the selected margin, the more difficult it is to establish non-inferiority. Given that the non-inferiority margin provides scientific credibility to the test, it should be selected carefully based on relevant evidence, as well as sound clinical, practical, and/or ethical considerations. Extant literature on non-inferiority testing provides various guidelines for the selection of the non-inferiority margin (e.g., Walker & Nowacki, 2011; Lakens, 2017). With respect to the SMART weight loss study, based on relevant evidence (Ross & Wing, 2016) and established guidelines (Mohr et al, 2012), the non-inferiority margin is selected to be 50% of the difference found between similar intervention approaches. In this case 50% of 11.24 lb is 5.62 lb. Hence, the non-inferiority of AI #1 in relation to AI #3 will be established if the data provide sufficient evidence to conclude that on average the advantage of AI #3 (in terms of weight loss by month 6) compared to AI #1 does not exceed 5.62 lb.

A related testing framework in randomized trials is that of *equivalence tests* (Seaman & Serlin, 1998; Goertzen & Cribbie, 2010; Walker & Nowacki, 2011; Lakens, 2017). In equivalence testing, one aims to test the hypothesis that a particular intervention is neither superior nor inferior to another intervention, subject to a pre-specified *equivalence margin*. The equivalence margin defines a range of values for which the expected outcomes are "close enough" to be considered equivalent. The equivalence of a new intervention is established when the data provide enough evidence to conclude that its efficacy is "close enough" (i.e., neither higher nor lower) to that of the active control intervention.

**Data Structure and Notation**

Consider a SMART design where individuals are randomized at the first stage to one of two initial intervention options, namely App (denoted by $a$) and App + Coaching (denoted by $ac$) as in Figure 2. Based on the individual's progress during the first intervention stage, s/he can be classified as either a responder ($R = 1$) or a non-responder ($R = 0$). At the second stage, responders are allowed to continue with the same initial intervention option, whereas non-responders are re-randomized to one of two second-stage options: modest augmentation (denoted by $m$) or vigorous augmentation (denoted by $v$). As can be seen in Figure 2 (also listed in Table 1), four AIs are embedded in this SMART: $d_1 = (a, a^R v^{1-R})$, $d_2 = (a, a^R m^{1-R})$, $d_3 = (ac, ac^R v^{1-R})$, and $d_4 = (ac, ac^R m^{1-R})$, where for example, $d_1$ is an AI that recommends offering intervention option $a$ initially to all target individuals, and then continue offering the same treatment for individuals who show early signs of response, while offering intervention option $v$ to those showing early signs of non-response. For clarity in comparing embedded AIs, we classify any pair of embedded AIs under comparison to be either (1) *distinct-path* (DP): those starting with different initial interventions (e.g., $\{d_1, d_3\}$ or $\{d_1, d_4\}$ or $\{d_2, d_3\}$ or $\{d_2, d_4\}$) or (2) *shared-path* (SP): those starting with the same initial intervention (e.g., $\{d_1, d_2\}$ or $\{d_3, d_4\}$) (Kidwell & Wahed, 2013).

Here, we assume a single continuous primary outcome $Y$ (e.g., the amount of weight loss) observed at the end of the trial. Let $T_1$ and $T_2$ generically denote the intervention options at stages 1 and 2. Then the observed data trajectory for the $i^{th}$ individual in a SMART is given by $(T_{1i}, R_i, T_{2i}, Y_i)$, $i = 1, \cdots, N$, where $N$ is the total number of individuals in the trial. Since the distribution of the primary outcome is indexed by the sequences of intervention options received, we write an individual's potential outcomes (Robins, 1997) as $Y_{T_1,T_2}$ where $T_1 \in \{a, ac\}; T_2 \in \{a, m, v\}$ if $T_1 = a$; and $T_2 \in \{ac, m, v\}$ if $T_1 = ac$. Then we assume

$$E(Y_{T_1,T_2}) = \mu_{T_1,T_2} \quad and \quad Var(Y_{T_1,T_2}) = \sigma^2, \tag{1}$$

which states that the mean outcome of all the individuals who were assigned the interventional sequence $\{T_1, T_2\}$ is $\mu_{T_1,T_2}$ and the corresponding variance is $\sigma^2$ (same for all the intervention sequences). Furthermore, the usual assumptions about potential outcomes (Robins, 1997) in a longitudinal setting, namely, (i) *consistency*, (ii) *no unmeasured confounding* and (iii) positivity are assumed to hold. Specifically, the *consistency* assumption states that the potential outcome under the observed treatment (sequence) and the observed outcome agree, i.e., $Y = Y_{T_1,T_2}$ if the observed treatment sequence is indeed $(T_1, T_2)$. The consistency assumption encompasses Rubin's (1980) stable unit treatment value assumption (SUTVA) that says each participant's potential outcome is not influenced by the treatment administered to the other participants (Chakraborty and Murphy, 2014). The *no unmeasured confounding* assumption states that treatment allocation is independent of future potential outcomes given the history; this is satisfied by design in case of a SMART (Murphy, 2005). Finally, the *positivity* assumption ensures that an individual has a positive probability of receiving any of the treatment sequences considered in the study. In other words, by positivity assumption, at least some individuals obtain each of the possible treatment sequences (AIs) feasible in the study.

For both the non-inferiority and the equivalence test procedures, the test statistic is based on the means of the embedded AIs. By design, the mean of an AI is a weighted average of primary outcomes of patients having treatment trajectories consistent with that AI (Oetting et al., 2011; Nahum-Shani et al., 2012a). Intuitively, a weighted average is essential as there is a structural imbalance between responders and non-responders. In the two-stage SMART considered here, non-responders are re-randomized at the second stage, but responders continue with the same treatment. Define the first-stage randomization probability in favor of intervention option $T_1$ as $\pi_{T_1}$, and the second-stage randomization probability for those who started with the first-stage option $T_1$, in favor of intervention option $T_2$ as $\pi_{T_1, T_2}$ (often these probabilities are 0.5). Note that, $\pi_{ac} = 1 - \pi_a$, $\pi_{a,v} = 1 - \pi_{a,m}$ and $\pi_{ac,v} = 1 - \pi_{ac,m}$. A responder, due to a single randomization at stage 1, is assigned to the treatment sequence $(T_1, T_1)$ with probability $\pi_{T_1}$ (0.5 in the case of the SMART weight loss study). On the other hand, due to two stages of randomization, a non-responder is assigned to a treatment sequence $(T_1, T_2)$ with probability $\pi_{T_1} \times \pi_{T_1, T_2}$ (0.5*0.5=0.25 in the case of the SMART weight loss study). Using the principles of *inverse probability weighting* (Robins, 1997; Nahum-Shani et al., 2012a), the weight used for the $i^{th}$ individual in any embedded AI $d$ is $\frac{1}{\pi_{T_{1i}}}$ for a responder ($\frac{1}{0.5} = 2$ in the case of responders in the SMART weight loss study) and $\frac{1}{\pi_{T_{1i}} \times \pi_{T_{1i}, T_{2i}}}$ for a non-responder ($\frac{1}{0.25} = 4$ in the case of the non-responders in the SMART weight loss study). In general, the weight can be expressed as $1/(\pi_{T_1} \times \pi_{T_{1i}, T_{2i}}^{1-R_i})$. Thus, the observed mean corresponding to the AI $d_1$ can be written as

$$\overline{Y}_{d_1} = \frac{\sum_{i=1}^{N} W_i^{d_1} Y_i}{\sum_{i=1}^{N} W_i^{d_1}}, \quad where \quad W_i^{d_1} = \frac{I\{T_{1i} = a, T_{2i} = a^{R_i} v^{1-R_i}\}}{\pi_a \times \pi_{a,v}^{1-R_i}}, \quad (2)$$

where $I\{T_{1i} = a, T_{2i} = a^{R_i} v^{1-R_i}\}$ is an indicator for whether (=1) or not (=0) individual $i$ is consistent with embedded AI $d_1$ (i.e., the individual was assigned to $a$ initially, and then continued with $a$ if s/he was a responder or assigned to $v$ if s/he was a non-responder). Similarly, we can define other weights $W_i^{d_2}, W_i^{d_3}$ and $W_i^{d_4}$ corresponding to the AIs $d_2, d_3$ and $d_4$, respectively. Define $\gamma_{T_1} \in \{\gamma_a, \gamma_{ac}\}$ as the response rate to the initial intervention options $T_1$, with respect to some pre-specified definition of response (e.g., in the weight loss SMART, the response is defined as losing at least 0.5 lb per week on average). Then the population mean of any embedded AI $d$ can be expressed as

$$\mu_d = E(\overline{Y}_d) = \gamma_{T_1}\mu_{T_1,T_1} + (1 - \gamma_{T_1})\mu_{T_1,T_2}, \tag{3}$$

and the variance of the observed (sample) AI mean as

$$Var(\hat{\mu}_d) = Var(\overline{Y}_d)$$

$$= \frac{1}{N}\left\{\frac{1-\gamma_{T_1}+\gamma_{T_1}\pi_{T_1,T_2}}{\pi_{T_1}\pi_{T_1,T_2}}\sigma^2 + \frac{\gamma_{T_1}(1-\gamma_{T_1}\pi_{T_1})}{\pi_{T_1}}\mu_{T_1,T_1}^2\right.$$

$$\left. + \frac{(1-\gamma_{T_1})(1-(1-\gamma_{T_1})\pi_{T_1}\pi_{T_1,T_2})}{\pi_{T_1}\pi_{T_1,T_2}}\mu_{T_1,T_2}^2 - 2\gamma_{T_1}(1-\gamma_{T_1})\mu_{T_1,T_1}\mu_{T_1,T_2}\right\}. \tag{4}$$

When $\pi_{T_1} = \pi_{T_1,T_2} = \frac{1}{2}$, then

$$Var(\hat{\mu}_d) = \frac{1}{N}\sigma_d^2$$

$$= \frac{1}{N}\{2(2-\gamma_{T_1})\sigma^2 + \gamma_{T_1}(2-\gamma_{T_1})\mu_{T_1,T_1}^2 + (3-2\gamma_{T_1}-\gamma_{T_1}^2)\mu_{T_1,T_2}^2 - 2\gamma_{T_1}(1-$$

$$\gamma_{T_1})\mu_{T_1,T_1}\mu_{T_1,T_2}\}.$$

$$\tag{5}$$

See Appendix A.1 and A.2 for a detailed derivation of the AI mean and corresponding variance, respectively. While developing the non-inferiority and equivalence test procedures in the context of a SMART, for simplicity, we consider $\pi_a = \pi_{ac} = \pi_{a,m} = \pi_{ac,m} = 0.5$. In other words, we assume that individuals are randomized with equal probability to first-stage intervention options, and non-responders as re-randomized with equal probability to second-stage intervention options.

**Non-Inferiority Test**

The objective of a classical non-inferiority test is to show that the efficacy of a new intervention, when compared to an active control, is not below a pre-specified non-inferiority margin. A non-inferiority test becomes a traditional one-sided superiority test when the non-inferiority margin is set to zero. In this section, we describe the non-inferiority test procedure in the context of a SMART. For completeness, we consider the non-inferiority test for both distinct-path and shared-path comparison of AIs. In general, for any two AIs $d_i$ and $d_j$ with corresponding means $\mu_{d_i}$ and $\mu_{d_j}$, the null and alternative hypotheses for the non-inferiority test can be written as

$$H_0: \mu_{d_i} - \mu_{d_j} \geq \theta \quad vs. \quad H_1: \mu_{d_i} - \mu_{d_j} < \theta, \tag{6}$$

where $\theta \ (> 0)$ is a pre-specified non-inferiority margin, and $j \neq i \in \{1, \cdots, 4\}$. This hypothesis tests that the average efficacy of $d_j$ (the 'new' AI) is not inferior to that of $d_i$ (the control, or 'standard of care' AI) with reference to the non-inferiority margin $\theta$. The unscaled test statistic is $\overline{Y}_{d_i} - \overline{Y}_{d_j}$. Intuitively, under the alternative, the difference in favor of the control AI ($d_i$) is smaller than $\theta$, while under the null, this difference is equal to, or higher than $\theta$. As noted earlier, the margin $\theta$ represents the maximum clinically acceptable difference in favor of the control AI ($d_i$) that an investigator is willing to accept in return for the secondary benefits (e.g., lower cost, burden and /or side effects) of the new AI ($d_j$). The choice of the non-inferiority margin depends on both statistical reasoning and clinical judgment (D'Agostino, Massaro, & Sullivan, 2003).

**Comparison of Distinct-Path AIs**

Let $\mu_{d_1}$ and $\mu_{d_3}$ be the means of the two distinct-path AIs, $d_1$ and $d_3$, respectively. Here we assume $d_3$ to be the active control (or standard of care) AI. The goal is to test whether $d_1$ is non-inferior in efficacy to $d_3$. From (6), the hypothesis of interest is

$$H_0: \mu_{d_3} - \mu_{d_1} \geq \theta \quad vs. \quad H_1: \mu_{d_3} - \mu_{d_1} < \theta.$$

The unscaled test statistic is $\overline{Y}_{d_3} - \overline{Y}_{d_1}$, with population mean $\mu_{d_3} - \mu_{d_1}$, and variance $\nu_{d_3 d_1}^{DP} = (\sigma_{d_3}^2 + \sigma_{d_1}^2)/N$ (see equation (5)), where

$$\sigma_{d_1}^2 = 2(2 - \gamma_a)\sigma^2 + \gamma_a(2 - \gamma_a)\mu_{a,a}^2 + (3 - 2\gamma_a - \gamma_a^2)\mu_{a,m}^2 - 2\gamma_a(1 - \gamma_a)\mu_{a,a}\mu_{a,m}, \text{ and}$$

$$\sigma_{d_3}^2 = 2(2 - \gamma_{ac})\sigma^2 + \gamma_{ac}(2 - \gamma_{ac})\mu_{ac,ac}^2 + (3 - 2\gamma_{ac} - \gamma_{ac}^2)\mu_{ac,v}^2 - 2\gamma_{ac}(1 - \gamma_{ac})\mu_{ac,ac}\mu_{ac,v}$$

$$(7).$$

Thus, under $H_0$, the large-sample distribution of the test statistic is

$$Z_{d_3 d_1} = \frac{(\overline{Y}_{d_3} - \overline{Y}_{d_1}) - (\mu_{d_3} - \mu_{d_1})}{\sqrt{var(\overline{Y}_{d_3} - \overline{Y}_{d_1})}} = \frac{\overline{Y}_{d_3} - \overline{Y}_{d_1} - \theta}{\sqrt{v_{d_3 d_1}^{DP}}} \to N(0,1).$$

Thus, we reject $H_0$ and conclude non-inferiority if $Z_{d_3 d_1} < -z_\alpha$ (one-sided test), where $\alpha$ is the Type-I error rate and $z_\alpha$ is the $(1 - \alpha)^{th}$ quantile of the standard normal distribution. Under a specified true mean difference $\delta$ such that $\mu_{d_3} - \mu_{d_1} = \delta$, the required sample size is

$$N = (z_\alpha + z_\beta)^2 \frac{(\sigma_{d_3}^2 + \sigma_{d_1}^2)}{(\theta - \delta)^2} = 2(z_\alpha + z_\beta)^2 / (\eta_{DP}(\theta) - \eta_{DP}(\delta))^2 = 2(z_\alpha + z_\beta)^2 / \eta_{DP}^2, \quad (8)$$

where $\beta$ is the Type II error rate, $\eta_{DP}(\theta) = \frac{\theta}{\sqrt{(\sigma_{d_3}^2 + \sigma_{d_1}^2)/2}}$ is the standardized margin,

$\eta_{DP}(\delta) = \frac{\delta}{\sqrt{(\sigma_{d_3}^2 + \sigma_{d_1}^2)/2}}$ is the standardized mean difference, and $\eta_{DP} = \eta_{DP}(\theta) - \eta_{DP}(\delta) =$

$\frac{\theta - \delta}{\sqrt{(\sigma_{d_3}^2 + \sigma_{d_1}^2)/2}}$ is the overall standardized effect size for the comparison of two distinct-path AIs,.

Furthermore, when $\delta = 0$, $\eta_{DP}$ boils down to the standardized margin. See Appendix A.3 for the derivation. Note that $N$ depends on the difference between the non-inferiority margin $\theta$ and the true mean difference $\delta$ rather than their individual values. Intuitively, $\theta - \delta$ represents the discrepancy between the acceptable difference in favor of the control AI ($d_3$) that one is willing to accept in return for the secondary benefits (e.g., lower cost and participant burden) of the alternative AI $(d_1)$, and the true mean difference in favor of $d_3$ (i.e., $\delta$). The smaller that discrepancy, the larger the total sample size ($N$) required to correctly establish non-inferiority with power 1- $\beta$, given the selected Type-I error rate $\alpha$. The quantity $\eta_{DP}$ is the standardized difference between $\theta$ and $\delta$.

**Comparison of Shared-Path AIs**

Next, we consider two shared-path AIs $d_3$ and $d_4$ with corresponding means $\mu_{d_3}$ and $\mu_{d_4}$. As before, consider $d_3$ to be the active control (or standard of care) AI. However, now the goal is to investigate whether $d_4$ is non-inferior to $d_3$. Notice that both AIs start with the same initial intervention option (app + coaching), however, while $d_3$ offers vigorous augmentation to non-responders in the form of text messaging and meal replacement, $d_4$ offers modest augmentation in the form of text messaging alone. As in the distinct-path case, the unscaled test statistic is $\overline{Y}_{d_3} - \overline{Y}_{d_4}$ with mean $\mu_{d_3} - \mu_{d_4}$. However, in this case, the variance of the unscaled test statistic will be different from that of the distinct-path AIs, because the set of responders to the initial intervention (app + coaching) is consistent with both $d_3$ and $d_4$, thereby inducing some correlation between the two AIs. The variance of $\overline{Y}_{d_3} - \overline{Y}_{d_4}$ is given by

$$v_{d_3 d_4}^{SP} = V(\hat{\mu}_{d_3}) + V(\hat{\mu}_{d_4}) - \frac{2}{N}\left[\frac{\gamma_{ac}}{\pi_{ac}}(\sigma^2 + \mu_{ac,ac}^2) - (\gamma_{ac}\mu_{ac,ac} + (1-\gamma_{ac})\mu_{ac,m}) \cdot \right.$$

$$(\gamma_{ac}\mu_{ac,ac} + (1-\gamma_{ac})\mu_{ac,v})]$$

$$= \frac{1}{N}(\sigma_{d_3}^2 + \sigma_{d_4}^2 - 2\sigma_{d_3 \times d_4}^2), \tag{9}$$

where $\frac{1}{N}\sigma_{d_3 \times d_4}^2 = cov(\hat{\mu}_{d_3}, \hat{\mu}_{d_4})$. See Appendix A.4 for the derivation. Therefore, under $H_0$, the large-sample distribution of the test statistic is

$$Z_{d_3 d_4} = \frac{(\overline{Y}_{d_3} - \overline{Y}_{d_4}) - (\mu_{d_3} - \mu_{d_4})}{\sqrt{var(\overline{Y}_{d_3} - \overline{Y}_{d_4})}} = \frac{\overline{Y}_{d_3} - \overline{Y}_{d_4} - \theta}{\sqrt{v_{d_3 d_4}^{SP}}} \to N(0,1).$$

Hence, we reject $H_0$ and conclude non-inferiority if $Z_{d_3 d_4} < -z_\alpha$ (one-sided). Similar to the distinct-path case, under a specified true mean difference $\delta$ such that $\mu_{d_3} - \mu_{d_4} = \delta$, the required sample size for the trial is

$$N = (z_\alpha + z_\beta)^2 \frac{\left(\sigma_{d_3}^2 + \sigma_{d_4}^2 - 2\sigma_{d_3 \times d_4}^2\right)}{(\theta - \delta)^2} \;\; = \;\; 2(z_\alpha + z_\beta)^2 / \left(\eta_{SP}(\theta) - \eta_{SP}(\delta)\right)^2 = 2(z_\alpha + z_\beta)^2 /$$

$$\eta_{SP}^2, \qquad\qquad (10)$$

where $\eta_{SP} = \eta_{SP}(\theta) - \eta_{SP}(\delta) = \frac{\theta - \delta}{\sqrt{\left(\sigma_{d_3}^2 + \sigma_{d_4}^2 - 2\sigma_{d_3 \times d_4}^2\right)/2}}$ is the standardized effect size for the comparison of two shared-path AIs. As in case of $\eta_{DP}$, the standardized effect size $\eta_{SP}$ is the difference between the standardized margin $\eta_{SP}(\theta) = \frac{\theta}{\sqrt{\left(\sigma_{d_3}^2 + \sigma_{d_4}^2 - 2\sigma_{d_3 \times d_4}^2\right)/2}}$ and the standardized mean difference $\eta_{SP}(\delta) = \frac{\delta}{\sqrt{\left(\sigma_{d_3}^2 + \sigma_{d_4}^2 - 2\sigma_{d_3 \times d_4}^2\right)/2}}$. Note that, in the shared-path

comparison of two AIs, only one response rate (e.g., in this case, that of the initial intervention

$ac$) is involved.

Note that in real trials, there is always some degree of attrition; hence in order to maintain power the sample size ($N$) needs to be inflated accordingly (in both distinct-path and shared-path scenarios).

**Standardized Effect Sizes**

It is important to note that even though we use the term standardized effect size here, the meaning of this effect size is different from Cohen's *d*, as it does not quantify the standardized difference between two means alone, but rather the standardized difference between the non-inferiority margin and the true difference between the two means. Hence, the benchmark values recommended by Cohen do not directly apply.

Furthermore, observe that for $\eta_{DP} = \eta_{SP}$, the required sample sizes (N) are same for comparison

of two distinct-path and two shared-path AIs. However, one should note that although the sample

sizes are same, they may refer to different contexts of non-inferiority tests with different non-

inferiority margin and/or different mean difference of AIs. For example, consider a SMART

wherein the relevant embedded AIs have the following variances and covariance: $\sigma_{d_1}^2 = 3, \sigma_{d_3}^2 = 5$ , $\sigma_{d_4}^2 = 4$ and $\sigma_{d_3 \times d_4}^2 = 0.5$. For distinct-path comparison of AIs $\{d_1, d_3\}$ with $\theta_{DP} = 1.4$ and $\delta_{DP} = 1$, the $\eta_{DP}$ is 0.2. On the other hand, for shared-path comparison of AIs $\{d_3, d_4\}$ with $\theta_{SP} = 2.8$ and $\delta_{SP} = 2.4$, the $\eta_{SP}$ is also 0.2. Thus, in both scenarios, the required sample size is 268. However, the key contextual parameters like the underlying standardized margin and standardized mean difference are different. One can claim that a particular SMART is powered for both distinct-path and shared-path comparisons only if it is acceptable to assume two different standardized margins for the two comparisons within the same SMART, which is unlikely the case in real life. Therefore, an investigator should first decide on the primary mode of comparison (distinct-path vs. shared-path) and determine the key context parameters (standardized margin, postulated standardized mean difference) before moving to the power/sample size calculation based on the standardized effect size derived from the above key context parameters. The same cautionary note applies to the equivalence tests discussed below.

**Equivalence Test**

As mentioned earlier, the term equivalence means that the efficacies of the two interventions under comparison are close to each other in a way that cannot be considered superior or inferior. Similar to the non-inferiority margin, here we consider an equivalence margin to define the closeness of two interventions. An equivalence test becomes a traditional two-sided superiority test when the equivalence margin is set to zero. An equivalence margin $(-\theta, \theta)$ is defined at the beginning of a trial such that any value in that range is clinically unimportant. After the trial, an equivalence is achieved if the estimated confidence interval of the treatment difference falls within $(-\theta, \theta)$ (Ebbutt & Frith, 1998). In general, for any two AIs $d_i$ and $d_j$ with corresponding means $\mu_{d_i}$ and $\mu_{d_j}$, the null and alternative hypotheses for the equivalence test are specified as

$$H_0: \mu_{d_i} - \mu_{d_j} \geq \theta \quad or \quad \mu_{d_i} - \mu_{d_j} \leq -\theta \quad vs. \quad H_1: -\theta < \mu_{d_i} - \mu_{d_j} < \theta, \quad (11)$$

where $\theta$ ($> 0$) is a pre-specified equivalence margin. The above test actually contains two separate sub-tests: (1) a non-inferiority test ($H_{01}: \mu_{d_i} - \mu_{d_j} \geq \theta \quad vs. \quad H_{11}: \theta < \mu_{d_i} - \mu_{d_j} < \theta$) and (2) a non-superiority test ($H_{02}: \mu_{d_i} - \mu_{d_j} \leq -\theta \quad vs. \quad H_{12}: \mu_{d_i} - \mu_{d_j} > -\theta$). In traditional RCTs, due to ease of use and recommendation from the United States Food and Drug Administration (FDA), a *two one-sided tests* (TOST) procedure is widely used for equivalence tests (Schuirmann, 1987; Lauzon & Caffo, 2009). Note that, in a TOST, two one-sided tests are considered, each of which having Type-I error rate $\alpha$. Even though in order to achieve equivalence, both sub-tests require to reject the corresponding null hypotheses, the overall Type-I error rate of the TOST procedure is still $\alpha$, because only one of the two null hypotheses corresponding to the two sub-tests can be true at any one point of time. We further illustrate this point using simulation studies; see results presented in Tables 6 and 7 below. However, it turns out that a TOST procedure is operationally identical to the procedure of declaring equivalence if the ordinary (1 - 2α) x100% confidence interval (CI) for the mean difference of two AIs is completely contained in the equivalence margin (-θ, θ) (Schuirmann, 1987). In this section, we develop an equivalence test procedure for SMART, separately for the distinct-path and the shared-path AIs.

**Comparison of Distinct-Path AIs**

First, we consider the non-inferiority sub-test of an equivalence test for distinct-path AIs. For the two distinct-path AIs $d_1$ and $d_3$ with respective means $\mu_{d_1}$ and $\mu_{d_3}$, we repeat the non-inferiority test procedure described in the previous section. The same test statistic $Z_{d_3 d_1}$ is used to test $H_{01}: \mu_{d_3} - \mu_{d_1} \geq \theta$ $vs.$ $H_{11}: \mu_{d_3} - \mu_{d_1} < \theta$. The null hypothesis $H_{01}$ will be rejected if $Z_{d_3 d_1} < -z_\alpha$.

For the non-superiority sub-test $H_{02}: \mu_{d_3} - \mu_{d_1} \leq -\theta$ $vs.$ $H_{12}: \mu_{d_3} - \mu_{d_1} > -\theta$, the large-sample distribution of the test statistic under $H_{02}$ is

$$Z^*_{d_3 d_1} = \frac{(\overline{Y}_{d_3} - \overline{Y}_{d_1}) - (\mu_{d_3} - \mu_{d_1})}{\sqrt{var(\overline{Y}_{d_3} - \overline{Y}_{d_1})}} = \frac{\overline{Y}_{d_3} - \overline{Y}_{d_1} + \theta}{\sqrt{v^{DP}_{d_3 d_1}}} \to N(0,1).$$

Notice that, $Z^*_{d_3 d_1}$ is different from $Z_{d_3 d_1}$ with respect to the sign of $\theta$ in their expressions. Reject $H_{02}$ and conclude non-superiority if $Z^*_{d_1 d_3} > z_\alpha$. Finally, we conclude equivalence when both tests reject the corresponding null hypotheses. Given a postulated true mean difference $\delta = \mu_{d_3} - \mu_{d_1} \neq 0$, there is no closed-form solution for the minimum required sample size ($N$). However, the power of the equivalence test is given by

$$\Phi\left(-z_\alpha + \frac{\theta - \delta}{\sqrt{v^{DP}_{d_3 d_1}}}\right) - \Phi\left(z_\alpha - \frac{\theta + \delta}{\sqrt{v^{DP}_{d_3 d_1}}}\right)$$

$$= \Phi\left(-z_\alpha + (\eta_{DP}(\theta) - \eta_{DP}(\delta))\sqrt{\frac{N}{2}}\right) - \Phi\left(z_\alpha - (\eta_{DP}(\theta) + \eta_{DP}(\delta))\sqrt{\frac{N}{2}}\right), \quad (12)$$

where the standardized margin $\eta_{DP}(\theta) = \frac{\theta}{\sqrt{(\sigma^2_{d_3} + \sigma^2_{d_1})/2}}$, the standardized mean difference $\eta_{DP}(\delta) = \frac{\delta}{\sqrt{(\sigma^2_{d_3} + \sigma^2_{d_1})/2}}$ and $\Phi(x)$ denotes the cumulative distribution function of a standard normal variable evaluated at a point $x$. See Appendix A.5 for the derivation. The formula in (12) can be used to estimate the power corresponding to a particular sample size ($N$) when all the involved parameters are known. Repeating the procedure to estimate the power corresponding to different $N$ leads to an iterative procedure to derive the required sample size corresponding to a desired power. Note that, unlike the comparison of distinct-path AIs in the non-inferiority setting, here the sample size based on expression (12) depends on the individual values of $\theta$ and $\delta$, as opposed to their difference $\theta - \delta$ alone.

In the special case when the postulated true mean difference $\delta = 0$, the minimum sample size has a closed form and is given by

$$N = \left(z_\alpha + z_{\beta/2}\right)^2 \frac{(\sigma_{d_3}^2 + \sigma_{d_1}^2)}{\theta^2} = \frac{2\left(z_\alpha + z_{\frac{\beta}{2}}\right)^2}{\eta_{DP}^2(\theta)}. \tag{13}$$

**Comparison of Shared-Path AIs**

In an equivalence test procedure to compare two shared-path AIs $d_3$ and $d_4$, as before, we first focus on the non-inferiority sub-test described above. The test statistic $Z_{d_3 d_4}$ is used to test $H_{01}$: $\mu_{d_3} - \mu_{d_4} \geq \theta$ $vs.$ $H_{11}$: $\mu_{d_3} - \mu_{d_4} < \theta$. The null hypothesis $H_{01}$ will be rejected if $Z_{d_3 d_4} < -z_\alpha$. For the non-superiority sub-test $H_{02}$: $\mu_{d_3} - \mu_{d_4} \leq -\theta$ $vs.$ $H_{12}$: $\mu_{d_3} - \mu_{d_4} > -\theta$, the large-sample distribution of the test statistic under $H_{02}$ is

$$Z_{d_3 d_4}^* = \frac{(\overline{Y}_{d_3} - \overline{Y}_{d_4}) - (\mu_{d_3} - \mu_{d_4})}{\sqrt{var(\overline{Y}_{d_3} - \overline{Y}_{d_4})}} = \frac{\overline{Y}_{d_3} - \overline{Y}_{d_4} + \theta}{\sqrt{v_{d_3 d_4}^{SP}}} \to N(0,1).$$

We reject $H_{02}$ and conclude non-superiority if $Z_{d_3 d_4}^* > z_\alpha$. As in the distinct-path scenario, we conclude equivalence when both sub-tests reject the corresponding null hypotheses. The power of the test for a given $N$ can be obtained from the expression (12) with $\eta_{DP}(\theta)$ and $\eta_{DP}(\delta)$ replaced by $\eta_{SP}(\theta)$ and $\eta_{SP}(\delta)$, respectively, where the standardized margin $\eta_{SP}(\theta) = \frac{\theta}{\sqrt{(\sigma_{d_3}^2 + \sigma_{d_4}^2 - 2\sigma_{d_3 \times d_4}^2)/2}}$ and the standardized mean difference $\eta_{SP}(\delta) = \frac{\delta}{\sqrt{(\sigma_{d_3}^2 + \sigma_{d_4}^2 - 2\sigma_{d_3 \times d_4}^2)/2}}$.

To estimate the minimum required sample size when the postulated true mean difference $\delta \neq 0$, one can compute the power for a series of $N$ until ones finds a suitable $N$ corresponding to a desired power.

Finally, in the special case of $\delta = 0$, the minimum sample size is given by

$$N = \left(z_\alpha + z_{\beta/2}\right)^2 \frac{\left(\sigma_{d_3}^2 + \sigma_{d_4}^2 - 2\sigma_{d_3 \times d_4}^2\right)}{\theta^2} = \frac{2\left(z_\alpha + z_{\frac{\beta}{2}}\right)^2}{\eta_{SP}^2(\theta)}. \tag{14}$$

## Simulation Studies

In this section, we discuss simulation studies conducted to: (a) investigate how the Monte Carlo (MC) power varies with the standardized effect size ($\eta$) and the sample size ($N$), when the margin ($\theta$) is fixed, by studying power curves; (b) empirically assess the performance of the sample size formulas in both non-inferiority and equivalence tests when the nominal power is fixed at 80%; (c) check the robustness of our sample size formula to the key working assumption of equal variance across various intervention sequences; and (d) illustrate the control of Type I error rate at or below the nominal 5% level under the TOST procedure employed in equivalence testing. The R-codes related to the simulation studies are available in Open Science Framework (*https://osf.io/hwae3/*).

### Simulation Design

The data generation process and the various required parameter values under all the simulation scenarios are given in the Appendix (section A.6 and Table 1).

For the first goal of studying power curves, we do not calculate $N$ from the sample size formulas, but rather set $N$ to specific values (e.g., $N = 100$, 200, 300 or 500). For illustration purpose, we set the non-inferiority/equivalence margin $\theta = 2$. Then, for varying standardized effect sizes, we generate 1000 simulated SMART datasets with the same size $N$. Next, for each of the 1000 datasets, we calculate the test statistic, and depending on its value, either reject or accept the null hypothesis. Finally, we calculate the MC power as the percentage of tests for which the null hypothesis is rejected. For example, if the null hypothesis is rejected in 810 out of the 1000 tests, then the MC power is 81%. In the power curves, the horizontal axis corresponds to standardized effect size and the vertical axis corresponds to estimated power.

For the second goal of assessing the adequacy of the sample size formulas, we first set the Type I error rate at 5%, the desired power at 80%, and the standardized effect size at a fixed value (e.g., 0.3). Next, we use these parameters to calculate the required sample size based on the relevant sample size formula, namely, equation (8), (10), (13) or (14), depending on whether the focus is on non-inferiority/equivalence testing and distinct-path/shared-path comparison. For example, in the case of a non-inferiority distinct-path comparison, the calculated N for a 0.3 standardized effect size with 80% power and 5% type I error rate is 138. Then, based on the calculated N (e.g., N=138) we generate 1000 simulated SMART datasets with the same size; thereafter for each of the 1000 datasets, we calculate the test statistic and either reject or accept the null hypothesis based on its value. Finally, as in the case of studying power curves, we calculate the MC power as the percentage of the tests for which the null hypothesis is rejected. An estimated MC power close to the nominal value of 80% indicates that the sample size formula is adequate.

The third goal of checking the robustness of our sample size formulas to the key working assumption of equal variance across various intervention sequences (see equation (1)) is operationally similar to the second goal described above. Here we calculate a robust version of MC power ($\widehat{power}_{robust}$) using 1000 simulated datasets following the same steps as above, but the data generation is slightly different. Specifically, in any given simulation scenario, we first determine the total required sample size $N$ of the SMART based on the relevant formula; then for each possible intervention sequence $\{T_1, T_2\}$ in that SMART, we generate $N_{T_1, T_2}$ number of observations on the primary outcome $Y$ from $N(\mu_{T_1,T_2}, \sigma^2_{T_1T_2})$ such that $\sum_{T_1,T_2} N_{T_1,T_2} = N$, where $\sigma_{T_1,T_2}$ is a random number generated from the uniform distribution $(\max(0.5, \sigma - 1), \ \sigma + 1)$, with $\sigma$ being the assumed common standard deviation across all the intervention sequences. Thus, even though our working assumption for the sample size formula takes the variances across the intervention sequences to be equal, the data here are generated under different variances. Hence, a value of the robust version of power ($\widehat{power}_{robust}$) close to the nominal power of 80% would establish the robustness of our sample size formulas.

For the fourth goal of illustrating the control of Type I error rate at or below the nominal 5% level under the TOST procedure employed in equivalence testing, we first determine the required sample sizes based on our formulas and then generate data when there is no true effect. Based on similar Monte Carlo technique as described above, we assess the rejection rates of the equivalence test. We also investigate additional sample sizes in the range 500-5000.

**Results**

In Figure 3, we present the power curves for both the distinct-path and the shared-path settings in non-inferiority and equivalence tests based on MC simulations. As expected, given a fixed standardized effect size ($\eta$), the power increases as the sample size increases from 100 to 500, for all the four scenarios. For a given $\eta$, the amount of increase in power is substantial for a change in the sample size from 100 to 200, compared to the change in the sample size from 200 to 300 or from 300 to 500. The power curve representation conveys a requirement of samples in the range of 100-200 in order to detect a standardized effect size in the range of 0.4 to 0.2, respectively. In case of equivalence tests, the power curves become gradually symmetric when the sample size increases from 100 to 500; however, for small trial sizes, the small number of samples in each experimental subgroup results in deviation from approximate normality of AI means, which in turn results in the slight asymmetry in the power curves.

In Table 2, we focus on a non-inferiority test involving the comparison of two distinct-path AIs $d_1$ and $d_3$, wherein $N$ represents the required sample size calculated from the relevant sample size formula, and $\widehat{power}$ is the corresponding MC power. Note that the highest standardized effect size $\eta_{DP}$ in Table 2 is 0.379 with corresponding sample size $N = 87$. We do not choose any higher standardized effect size in order to avoid small sample sizes (say, less than 80), because a SMART design with fewer than 80 samples may end up with very few individuals for some intervention sequences. In Table 2, the $\widehat{power}$ values range from 0.79 to 0.84. The required sample size $N$ is a decreasing function of $\eta_{DP}$, as expected. The value of the margin is also varied in our study, $\theta \in \{2.5, 3.0\}$. A smaller value of $\theta$ results in a smaller standardized effect size, which in turn increases the required sample size.

Similarly, in Table 3, we focus on a non-inferiority test involving the comparison of two shared-path AIs $d_4$ and $d_3$. Here we consider only one (shared) response rate $\gamma_{ac}$. In Table 3, the $\widehat{power}$ values range from 0.77 to 0.84. As expected, the required sample size $N$ is a decreasing function of the corresponding standardized effect size.

The last columns in both Tables 2 and 3 show the estimated robust power ($\widehat{power}_{robust}$) values, where the data generation violates the working assumption of equal variance across intervention sequences. These values range between 0.79 and 0.82 in Table 2 and between 0.76 and 0.84 in Table 3, and are either same or close to the corresponding $\widehat{power}$ values, thereby establishing the robustness of our proposed sample size formulas.

As explained earlier, in an equivalence test, there is no closed-form expression of the required sample size in general; however, in the special case when $\delta = 0$, the sample size can be written in a closed form. In Tables 4 and 5, we present simulation results from equivalence tests in distinct-path and shared-path comparisons of two AIs, respectively. Here we assume $\delta = \delta_{DP} = \delta_{SP} = 0$ and $\gamma_a = \gamma_{ac} = \gamma$. From Tables 3 and 4, clearly the required sample size $N$ is a decreasing function of $\gamma$ for both the distinct-path and the shared-path comparisons. In Table 4 (distinct-path comparison), the $\widehat{power}$ values range from 0.82 to 0.84; likewise, in Table 5 (shared-path comparison), the $\widehat{power}$ values range from 0.83 to 0.86.

In Tables 6 and 7, we have verified that the Type-I error rate is controlled at or below 5% (for nominal $\alpha = 0.05$) when there is no effect. In both tables, the respective sample sizes 244 and 175 in the first rows are based on the sample size formula for the equivalence test when $\delta = \delta_{DP} = \delta_{SP} = 0$. Other values of N, 500-5000, are provided when the standardized effect sizes 0.265 and 0.313 are fixed in Tables 6 and 7, respectively. The estimate Type-I error rates in both the tables establish our claim that the Type I error rate is controlled at or below 5% when there is no effect even with increasing sample sizes.

**Discussion**

Non-inferiority and equivalence tests are widely used in traditional randomized controlled trials (RCTs) to establish the non-inferiority or the equivalence of one intervention, which has some secondary benefit (e.g. less costly or less burdensome), over the other. Blackwelder (1982) gave a detailed discussion on the inappropriate use of the usual null hypothesis to show that a new therapy was as effective as a standard therapy. The roles of the null and alternative hypotheses (hence the Type-I and Type-II error rates) are reversed from traditional superiority test to a non-inferiority (and equivalence) test. An investigator should take note of this reversal when interpreting results from a non-inferiority/equivalence test.

We motivated the need for non-inferiority and equivalence tests for SMARTs by considering an example from the SMART weight loss study. We have developed data analytic methods and sample size formulas for SMART studies with primary scientific questions concerning the non-inferiority or equivalence of one adaptive intervention (AI) over another. The simulation studies demonstrate the adequacy of the sample size formulas. The estimated power (based on MC simulations) came out to be very close to the nominal value of 80% in all scenarios. The developed methodologies can handle situations in which the AIs under comparison begin with different intervention options (i.e., distinct-path) and also in which they begin with the same intervention option (i.e., shared-path). We illustrate the usefulness of our methodology in analyzing SMART data to conclude non-inferiority/equivalence among AIs through simulated data analyses in the Appendix (see section A.7 and Appendix Table 2).

The simulation studies also demonstrate the practical feasibility of testing non-inferiority and equivalence with data arising from SMARTs, given the realistic sample sizes needed to achieve adequate power. SMARTs enrolling 200-700 participants are highly common. For example, both the Adapt RBT SMART for developing an adaptive reinforcement-based treatment for pregnant women who are drug dependent (H. Jones, PI; NIH/NIDA R01DA014979; see Lei, Nahum-Shani, Lynch, Oslin, & Murphy, 2012) and the ExTENd SMART for extending treatment effectiveness of naltrexone (D. Oslin, PI, NIH/NIAAA; NIAAAOSL01485; see Nahum-Shani et al., 2017) enrolled about 300 participants, the ENGAGE SMART study for engaging cocaine-dependent individuals in treatment services (McKay et al., 2015) enrolled 500 participants, and the PLUTO SMART (Fu et al., 2017) for lung cancer screening and tobacco cessation (PLUTO) enrolled 1,000 participants. In this light, SMART with sample size in the range 50-200 may be considered a small SMART.

The methods described in the current manuscript are motivated by various SMART studies,

including the SMART Weight Loss study that we have used throughout for illustration. Given resource limitations, many SMART studies (see Lei et al., 2012) seek to inform the construction of stepped-care adaptive interventions (Sobell & Sobell, 2000), whereby minimal support (consisting of relatively low burden and/or low-cost intervention options) is offered initially, and more burdensome/costly interventions are reserved for those who need it most (i.e., those who do not respond adequately to minimal support). In this context, testing non-inferiority or equivalence of one embedded AI (e.g., the less costly/burdensome) over another (e.g., an AI that represents a more costly/burdensome standard of care) is logical and practically meaningful. In fact, in the current era of fast technological advancements, there is increasing interest in developing stepped-care behavioral interventions that use mHealth tools as minimal support. mHealth interventions capitalize on mobile and/or wireless devices to deliver behavioral interventions in a relatively low cost, low burden, and accessible manner (Kumar et al., 2013). The use of SMART studies to investigate the optimal way to integrate mHealth tools with more traditional intervention components in order to achieve more cost-effective and scalable intervention programs often necessitates non-inferiority and equivalence testing frameworks. For example, the justification for initiating treatment by offering a new mHealth tool (e.g., weight loss app) instead of a more costly/burdensome traditional intervention component (e.g., coaching or in-person sessions) requires demonstrating that the weight loss generated by an AI that integrates the new mHealth tool is only marginally lower than the more costly/burdensome traditional approach.

Whether the SMART should be sized for non-inferiority or equivalence depends on the scientific questions motivating the trial. Non-inferiority is used when the primary objective is to demonstrate that the novel approach (e.g., an AI that integrates low-cost or low-burden components such as mHealth tools) is not worse than an active control intervention (e.g., an AI that offers more costly/burdensome components throughout). In this case, the novel intervention is expected to be only marginally less beneficial (and possibly even more beneficial) than the active control intervention. Equivalence is used when the primary objective is to demonstrate that the novel intervention is neither worse nor better than the active control. Equivalency designs are rarely used in randomized trials aiming to construct or evaluate less costly/burdensome behavioral interventions because their typical objective is to show that the effectiveness of a new intervention is only marginally lower (i.e., non-inferior), as opposed to similar (i.e., equivalence) compared to the standard of care (Greene, Morland, Durkalski, & Frueh, 2008).

A SMART can have one (or multiple) primary hypothesis and multiple secondary hypotheses (*Almirall et al., 2014*). In an experimental trial, an investigator determines the required sample size based on the primary hypothesis (*Piantadosi, 2005*). However, s/he may have other secondary hypotheses which justify the inclusion of adaptive interventions (say, AI #2, AI # 4) other than main adaptive interventions of interest (say for distinct-path comparison: AI #1 vs. AI #3). However, not all the secondary hypotheses are necessarily adequately powered (*Murphy, 2005*). Nonetheless, these secondary or exploratory aims are important in advancing science as it can facilitate preliminary insights to inform future studies or to generate new hypotheses.

From an investigator's point of view, an advantage of the developed methods is that the required sample size can be calculated only by specifying the standardized effect size ( as a function of standardized margin and the standardized mean difference), without having to specify additional parameters that are often difficult to elicit (e.g., variances). To make these methods accessible to investigators, we have developed freely available online tools (https://osf.io/z7av5/) using R statistical software. These tools allow investigators to plan sample size for SMART studies aiming to test the non-inferiority or equivalence of one embedded AI compared to another without requiring any programming knowledge. They also allow investigators to analyze data from completed SMART studies to answer questions about non-inferiority and equivalence.

Our entire methodology of non-inferiority/equivalence testing as presented in this article relies on the classical frequentist paradigm, which concludes non-inferiority/equivalence based on a pre-specified threshold (say, 0.05) of the corresponding p-values. To overcome this over-dependence on the p-value cutoff in hypothesis testing, many researchers prefer a Bayesian analysis, and in particular, a quantification of evidence in favor of a hypothesis through a quantity called *Bayes Factor*. A full-fledged Bayesian analysis that needs to specify prior distributions for parameters under both the null and alternative hypotheses is beyond the scope of the current article. However, there exists a simple way to convert an observed p-value to construct an upper bound of the Bayes Factor; this bound is given by $\frac{1}{-e\,p\log(p)}$, where $p$ denotes the p-value (Benjamin and Berger, 2016; Bayarri et al., 2016). The above upper bound is the largest possible value of the Bayes Factor over any (reasonable) choice of the prior distribution for the alternative hypothesis, and thus represents the maximum possible evidence in favor of the alternative hypothesis present in the data. Being a simple function of the p-value, it is easy to calculate. Also, it is an acceptable metric to both Bayesians and frequentists. While illustrating our non-inferiority/equivalence testing methodology through some simulated dataset in the Appendix, we also report the estimated upper bounds of the Bayes Factor in addition to the usual p-values (see Appendix section A.7 and Appendix Table 2).

**Limitations and Directions for Future Research**

The described methods in the current manuscript are based on the so-called restricted SMART wherein only one group of participants (e.g., non-responders to the first-stage intervention) are re-randomized at the second stage. In principle, the developed methods, with some modification, can also be applied to other types of SMART designs discussed in the introduction. For example, in a SMART, when both the responders and non-responders from the first stage are re-randomized at the second stage, the mean and the variance formula of an AI can be derived from the same approach as in the Appendix, even though we have not explicitly done that. The rest of the developed procedures remain the same in the context of non-inferiority/equivalence tests.

In the current article, we have described the non-inferiority and equivalence testing framework with respect to continuous outcomes only. However, the same methodology can be easily extended to binary outcomes, following the work of Ghosh, Cheung, and Chakraborty (2016). For time-to-event outcomes, some additional work may be necessary to develop non-inferiority/equivalence versions of the weighted log-rank tests considered by Kidwell and Wahed (2013). Some work can be done to extend the proposed methodology for longitudinal outcome data (to incorporate within-person correlation) and for clustered SMARTs (to incorporate intraclass correlation).

The non-inferiority/equivalence margin $\theta$ is a key concept that formalizes the notion of "sufficiently close" in such trials. However, setting an appropriate value of $\theta$ remains a challenge. A margin that is too narrow may lead to unnecessarily large sample size requirements, whereas a margin that is too wide may end up establishing an equivalence for two interventions that are in fact substantially different. Information from recent clinical trials involving the standard treatment can provide useful guidelines for choosing an appropriate value of $\theta$ (D'Agostino et al., 2003). Further discussion of the selection of non-inferiority margin in an RCT can be found in Chow and Song (2016). In the context of SMARTs, if the treatment sequences consist of marketed drugs, data from the corresponding phase-III trials can be used to choose a value of $\theta$. More in-depth thought about how to optimally set the value of $\theta$ is necessary in the SMART design context; we view this as an important future work.

In this article, we have not considered a policy evaluation approach to find an optimal AI (Zhao et al., 2012, 2015a,b; Zhang et al., 2012; Logan et al., 2017; Laber et al., 2018). In particular, Logan et al. (2017) use *Bayesian Additive Regression Trees (BART)* to model the conditional mean function of the outcome and to inform identification of an individualized treatment rule (ITR) and subsequently proposed a method to find the optimal ITR. A policy evaluation approach in the context of non-inferiority/equivalence tests will be exciting future work. Specifically, it would be interesting to find an optimal AI based on the individuals' treatment and other covariates' information among a finite set of non-inferior (equivalent) AIs.

### Acknowledgements

Bibhas Chakraborty acknowledges support from the Duke-NUS Medical School, National University of Singapore, and an AcRF Tier 2 grant (MOE2015-T2-2-056) from the Ministry of Education, Singapore. Bonnie Spring and Inbal Nahum-Shani acknowledge that support for the Smart Weight Loss Management study was provided by National Institutes of Health grant R01DK108678. Inbal Nahum Shani also acknowledges support from the National Institutes of Health grants R01 DA039901 and P50 DA039838.

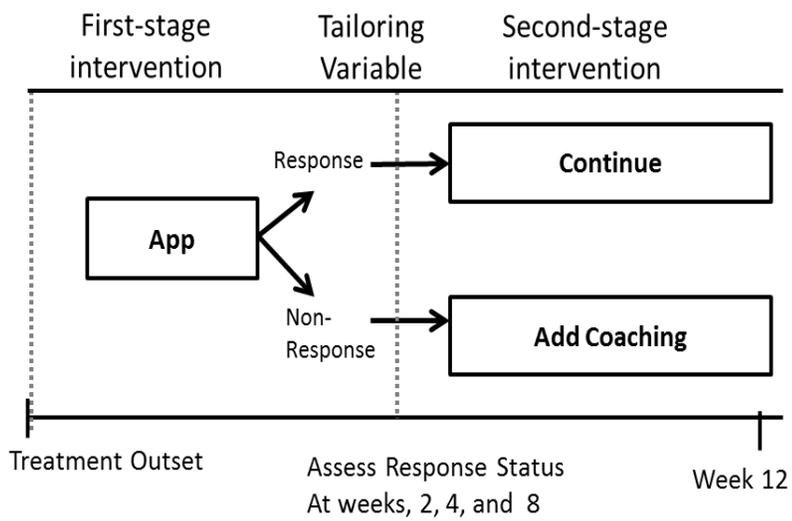

**App** → Mobile Application

Figure 1: An example adaptive intervention for promoting weight loss among obese/overweight adults

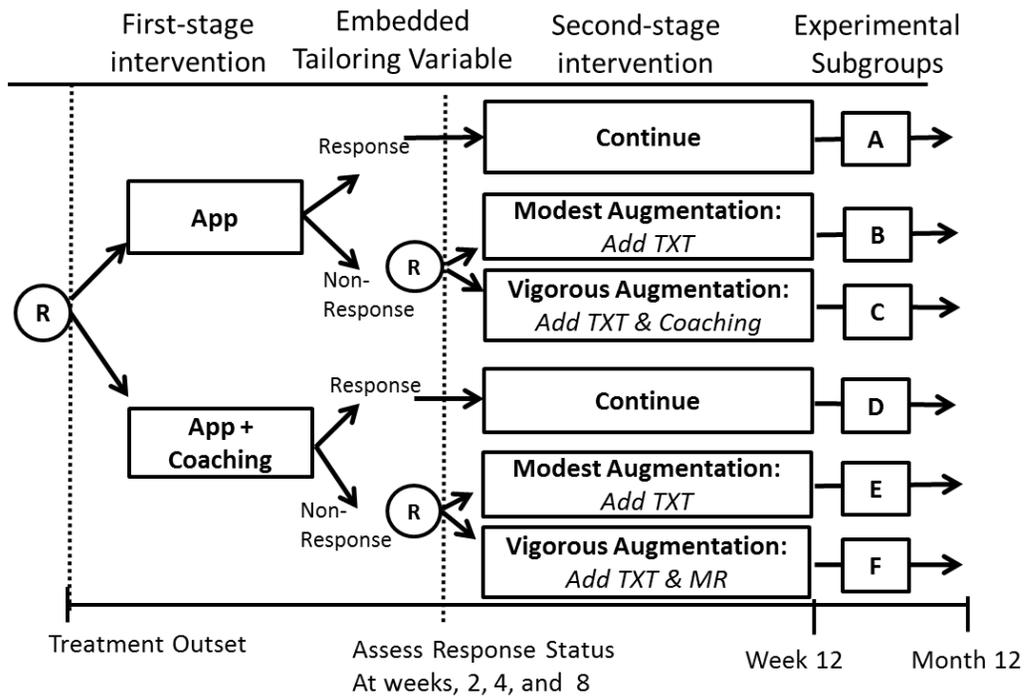

| First-stage intervention | Embedded Tailoring Variable | Second-stage intervention | Experimental Subgroups |
|---|---|---|---|

**App** → Mobile Application
**MR** → Meal Replacement
**TXT** → Supportive Text Messaging

Figure 2: The SMART Weight Loss Management Study

**Table 1.** Adaptive Interventions Embedded in the Weight Loss SMART Study

| AIs | Initial treatment | Subsequent tactic (treatment) for non-responders (R=0) | Subsequent tactic for responders (R=1) | Subgroups in Figure 1 | Stage-specific treatments: $(T_1, T_2)$ |
|---|---|---|---|---|---|
| 1 | App (a) | Vigorous Augmentation (v): add supportive text messaging and coaching | | A+C | $(a, a^R v^{1-R})$ |
| 2 | App (a) | Modest Augmentation (m): add supportive text messaging | Continue first-line intervention | A+B | $(a, a^R m^{1-R})$ |
| 3 | App + coaching (ac) | Vigorous Augmentation (v): add supportive text messaging and meal replacement | | D+F | $(ac, ac^R v^{1-R})$ |
| 4 | App + coaching (ac) | Modest Augmentation (m): add supportive text messaging | | D+E | $(ac, ac^R m^{1-R})$ |

Table 2. Results of the non-inferiority test involving the distinct-path AIs $\{d_1, d_3\}$.

| $(\gamma_a, \gamma_{ac})$ | $\theta$ | $\eta_{DP}$ | $N$ | $\widehat{power}$ | $\widehat{power}_{robust}$ |
|---|---|---|---|---|---|
| (0.30, 0.50) | 3.0 | 0.379 | 87 | 0.82 | 0.81 |
| (0.30, 0.45) | 3.0 | 0.371 | 90 | 0.80 | 0.80 |
| (0.30, 0.40) | 3.0 | 0.362 | 95 | 0.82 | 0.82 |
| (0.30, 0.35) | 3.0 | 0.354 | 99 | 0.82 | 0.81 |
| (0.30, 0.30) | 3.0 | 0.347 | 103 | 0.79 | 0.81 |
| (0.30, 0.50) | 2.5 | 0.251 | 197 | 0.80 | 0.79 |
| (0.30, 0.45) | 2.5 | 0.243 | 210 | 0.81 | 0.81 |
| (0.30, 0.40) | 2.5 | 0.236 | 223 | 0.80 | 0.80 |
| (0.30, 0.35) | 2.5 | 0.230 | 234 | 0.79 | 0.81 |
| (0.30, 0.30) | 2.5 | 0.223 | 249 | 0.84 | 0.82 |

Note. $\theta$ is the non-inferiority margin. $\gamma_a$ and $\gamma_{ac}$ are the response rates of initial treatments $a$ and $ac$, corresponding to the AIs $d_1$ and $d_3$ respectively; $\eta_{DP}$ is the standardized effect size. The required sample size $N$ is calculated based on equation (8) and the estimated power ($\widehat{power}$ and $\widehat{power}_{robust}$) is based on 1000 Monte Carlo simulations.

Table 3. Results of the non-inferiority test involving the shared-path AIs $\{d_4, d_3\}$

| $\gamma_{ac}$ | $\theta$ | $\eta_{SP}$ | $N$ | $\widehat{power}$ | $\widehat{power}_{robust}$ |
|---|---|---|---|---|---|
| 0.50 | 3.0 | 0.384 | 84 | 0.79 | 0.78 |
| 0.45 | 3.0 | 0.345 | 104 | 0.78 | 0.78 |
| 0.40 | 3.0 | 0.312 | 128 | 0.81 | 0.79 |
| 0.35 | 3.0 | 0.281 | 157 | 0.84 | 0.84 |
| 0.30 | 3.0 | 0.254 | 192 | 0.80 | 0.79 |
| 0.50 | 2.5 | 0.252 | 195 | 0.77 | 0.76 |
| 0.45 | 2.5 | 0.215 | 268 | 0.79 | 0.79 |
| 0.40 | 2.5 | 0.184 | 366 | 0.79 | 0.79 |
| 0.35 | 2.5 | 0.157 | 502 | 0.83 | 0.81 |
| 0.30 | 2.5 | 0.130 | 732 | 0.82 | 0.82 |

Note. $\theta$ is the non-inferiority margin.; $\gamma_a$ is the response rate of the initial treatment $a$ that corresponds to both AIs $d_4$ and $d_3$ (the other initial response rate $\gamma_{ac}$ is set at 0.5); $\eta_{SP}$ is the standardized effect size. The required sample size $N$ is calculated based on equation (10) and the estimated power ($\widehat{power}$ and $\widehat{power}_{robust}$) is based on 1000 Monte Carlo simulations.

Table 4. Results of the equivalence test involving the distinct-path AIs $\{d_1, d_3\}$ with $\delta = \mu_{d_3} - \mu_{d_1} = 0$

| ( $\gamma_a$ , $\gamma_{ac}$ ) | $\theta$ | $\eta_{DP}$ | $N$ | $\widehat{power}$ |
|---|---|---|---|---|
| (0.50, 0.50) | 2.0 | 0.265 | 244 | 0.83 |
| (0.45, 0.45) | 2.0 | 0.259 | 256 | 0.83 |
| (0.40, 0.40) | 2.0 | 0.254 | 266 | 0.82 |
| (0.35, 0.35) | 2.0 | 0.249 | 277 | 0.82 |
| (0.30, 0.30) | 2.0 | 0.244 | 288 | 0.84 |

Note. $\theta$ is the equivalence margin; $\gamma_a$ and $\gamma_{ac}$ are the response rates of initial treatments $a$ and $ac$, corresponding to the AIs $d_1$ and $d_3$ respectively; $\eta_{DP}$ is the standardized effect size. The required sample size $N$ is calculated based on equation (13) and the estimated power ($\widehat{power}$) is based on 1000 Monte Carlo simulations.

Table 5: Results of the equivalence test involving the shared-path AIs $\{d_4, d_3\}$ with $\delta = \mu_{d_3} - \mu_{d_4} = 0$

| $\gamma_{ac}$ | $\theta$ | $\eta_{SP}$ | $N$ | $\widehat{power}$ |
|---|---|---|---|---|
| 0.50 | 2.0 | 0.307 | 182 | 0.83 |
| 0.45 | 2.0 | 0.293 | 200 | 0.85 |
| 0.40 | 2.0 | 0.280 | 219 | 0.86 |
| 0.35 | 2.0 | 0.269 | 237 | 0.86 |
| 0.30 | 2.0 | 0.258 | 258 | 0.85 |

Note. $\theta$ is the equivalence margin; $\gamma_{ac}$ is the response rate of the initial treatment $ac$ that corresponds to both AIs $d_4$ and $d_3$ ($\gamma_a$ is the response rate to the other initial treatment $a$, and is set to be equal to $\gamma_{ac}$ while simulating the equivalence trials); $\eta_{SP}$ is the standardized effect size. The required sample size $N$ is calculated based on equation (14) and the estimated power ($\widehat{power}$) is based on 1000 Monte Carlo simulations.

Table 6. Results of the equivalence test when there is no effect involving the distinct-path AIs $\{d_1, d_3\}$

| $(\gamma_a , \gamma_{ac})$ | $\theta$ | $\eta_{DP}$ | $N$ | $Type-\widehat{I\ error\ rate}$ |
|---|---|---|---|---|
| | 2.0 | | 244 | 0.038 |
| | 2.0 | | 500 | 0.037 |
| (0.45, 0.45) | 2.0 | 0.265 | 1000 | 0.033 |
| | 2.0 | | 2000 | 0.043 |
| | 2.0 | | 5000 | 0.043 |

Note. $\theta$ is the equivalence margin; $\gamma_a$ and $\gamma_{ac}$ are the response rates of initial treatments $a$ and $ac$, corresponding to the AIs $d_1$ and $d_3$ respectively; $\eta_{DP}$ is the standardized effect size. The required sample size $N = 244$ is calculated based on equation (13) and the estimated Type-I error rate is based on 1000 Monte Carlo simulations. Other values of N, 500-5000, are provided to check that the Type-I error rate is controlled at 5% when there is no effect. Here the $\delta_{DP}$ is fixed at 2 ensuring there is no effect with respect to the equivalence margin $\theta = 2$.

Table 7: Results of the equivalence test when there is no effect involving the shared-path AIs $\{d_4, d_3\}$

| $\gamma_{ac}$ | $\theta$ | $\eta_{SP}$ | $N$ | $Type-\widehat{I\ error\ rate}$ |
|---|---|---|---|---|
| | 2.0 | | 175 | 0.066 |
| | 2.0 | | 500 | 0.039 |
| 0.45 | 2.0 | 0.313 | 1000 | 0.041 |
| | 2.0 | | 2000 | 0.055 |
| | 2.0 | | 5000 | 0.051 |

Note. $\theta$ is the equivalence margin; $\gamma_{ac}$ is the response rate of the initial treatment $ac$ that corresponds to both AIs $d_4$ and $d_3$ ($\gamma_a$ is the response rate to the other initial treatment $a$, and is set to be equal to $\gamma_{ac}$ while simulating the equivalence trials); $\eta_{SP}$ is the standardized effect size. The required sample size $N = 175$ is calculated based on equation (14) and the estimated Type-I error rate is based on 1000 Monte Carlo simulations. Other values of N, 500-5000, are provided to check whether the Type-I error rate is controlled at 5% when there is no effect. Here the $\delta_{SP}$ is fixed at 1.99 ensuring there is no effect with respect to the equivalence margin $\theta = 2$.

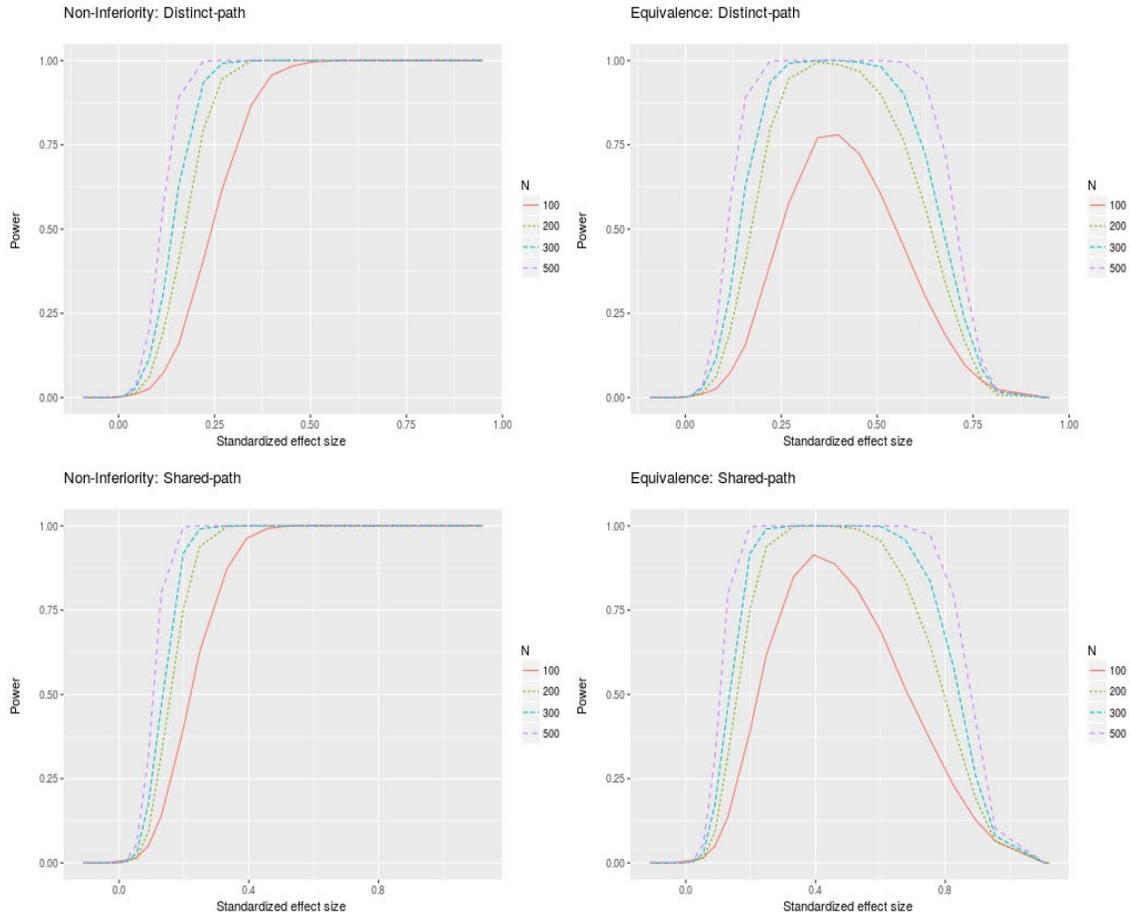

Figure 3: Power Curves for non-inferiority and equivalence tests for comparing distinct-path and shared-path adaptive interventions. For a given $N$ (100, 200, 300 or 500) and a standardized effect size (for distinct-path: $(\eta_{DP}(\theta) - \eta_{DP}(\delta)$ and for shared-path: $(\eta_{SP}(\theta) - \eta_{SP}(\delta))$, the power is based on 1000 Monte Carlo simulations.

## A   Appendix

### A.1   Derivation of Mean of an Adaptive Intervention

We will use $Y$ as the primary outcome instead of $Y_{T_1 T_2}$ to refer an individual's primary outcome with respect to the entire study. Note that the response indicator $R = I(T_1 = T_2)$; and, $\sum_{i=1}^{N} W_i^{d_1} = N$, when empirical randomization probabilities are exactly same as the corresponding theoretical probabilities. From (2), using the independence among the different patients in the study, we have

$$\mu_{d_1} = E(\overline{Y}_{d_1}) = E[W^{d_1} Y] = E\left\{ E\left[ \frac{I\{T_1 = a, T_2 = a^R v^{1-R}\}}{\pi_a \times \pi_{a,v}^{1-R}} Y \Big| T_1, T_2 \right] \right\}$$

$$= P(T_1 = a, T_2 = a) E\left[ \frac{I\{T_1 = a, T_2 = a\}}{\pi_a} Y \Big| T_1 = a, T_2 = a \right]$$

$$+ P(T_1 = a, T_2 = v) E\left[ \frac{I\{T_1 = a, T_2 = v\}}{\pi_a \times \pi_{a,v}} Y \Big| T_1 = a, T_2 = v \right] + 0,$$

(the   last   term:   if   $T_1 =$ ac   then   it   contributes   0   due   the   indicator   function)

$$= P(T_1 = a) P(T_2 = a | T_1 = a) E\left[ \frac{1}{\pi_a} Y \Big| T_1 = a, T_2 = a \right]$$

$$+ P(T_1 = a) P(T_2 = v | T_1 = a) E\left[ \frac{1}{\pi_a \times \pi_{a,v}} Y \Big| T_1 = a, T_2 = v \right]$$

$$= \pi_a \gamma_a \frac{1}{\pi_a} \mu_{a,a} + \pi_a (1 - \gamma_a) \pi_{a,v} \frac{1}{\pi_a \pi_{a,v}} \mu_{a,v}$$

$$= \gamma_a \mu_{a,a} + (1 - \gamma_a) \mu_{a,v}. \tag{15}$$

Similarly,   $\mu_{d_2} = \gamma_a \mu_{a,a} + (1 - \gamma_a) \mu_{a,m}$ ;   $\mu_{d_3} = \gamma_{a,c} \mu_{ac,ac} + (1 - \gamma_{ac}) \mu_{ac,v}$   and   $\mu_{d_4} = \gamma_{ac} \mu_{ac,ac} + (1 - \gamma_{ac}) \mu_{ac,m}$.

### A.2   Derivation of Variance of the Mean of an Adaptive Intervention

$$Var(\hat{\mu}_{d_1}) = \frac{1}{N} Var[W^{d_1}Y], \qquad (using \quad independent \quad observations)$$

$$= \frac{1}{N}\left\{Var\left(\frac{I\{T_1=a, T_2=a^R v^{1-R}\}}{\pi_a \times \pi_{a,v}^{1-R}}Y\right)\right\}$$

$$= \frac{1}{N}\left\{E\left[\frac{I\{T_1=a, T_2=a^R v^{1-R}\}}{(\pi_a \times \pi_{a,v}^{1-R})^2}Y^2\right] - E^2\left[\frac{I\{T_1=a, T_2=a^R v^{1-R}\}}{\pi_a \times \pi_{a,v}^{1-R}}Y\right]\right\}. \tag{16}$$

Note that, from (15), we have the expected mean of the AI $d_1$,

$$E\left[\frac{I\{T_1=a, T_2=a^R v^{1-R}\}}{\pi_a \times \pi_{a,v}^{1-R}}Y\right] = \gamma_A \mu_{a,a} + (1-\gamma_a)\mu_{a,v}. \tag{17}$$

Now, the first part of (16),

$$E\left[\frac{I\{T_1=a, T_2=a^R v^{1-R}\}}{(\pi_a \times \pi_{a,v}^{1-R})^2}Y^2\right]$$

$$= P(T_1=a, T_2=a)E\left[\frac{I\{T_1=a, T_2=a\}}{\pi_a^2}Y^2 | T_1=a, T_2=a\right]$$

$$+ P(T_1=a, T_2=v)E\left[\frac{I\{T_1=a, T_2=v\}}{\pi_a^2 \times \pi_{a,v}^2}Y^2 | T_1=a, T_2=v\right] + 0,$$

$$(the \quad last \quad term: \quad if \quad T_1 =$$

$$ac \quad then \quad it \quad contributes \quad 0 \quad due \quad the \quad indicator \quad function)$$

$$= P(T_1=a)P(T_2=a|T_1=a)E\left[\frac{1}{\pi_a^2}Y^2 | T_1=a, T_2=a\right]$$

$$+ P(T_1=a)P(T_2=v|T_1=a)E\left[\frac{1}{\pi_a^2 \times \pi_{a,v}^2}Y^2 | T_1=a, T_2=v\right]$$

$$= \pi_a \gamma_a \frac{1}{\pi_a^2}(\sigma^2 + \mu_{a,a}^2) + \pi_a(1-\gamma_a)\pi_{a,v}\frac{1}{\pi_a^2 \pi_{a,v}^2}(\sigma^2 + \mu_{a,v}^2)$$

$$= \frac{\gamma_a}{\pi_a}(\sigma^2 + \mu_{a,a}^2) + \frac{(1-\gamma_a)}{\pi_a \pi_{a,v}}(\sigma^2 + \mu_{a,v}^2)$$

$$= \frac{1-\gamma_a+\gamma_a\pi_{a,v}}{\pi_a \pi_{a,v}}\sigma^2 + \frac{\gamma_a}{\pi_a}\mu_{a,a}^2 + \frac{(1-\gamma_a)}{\pi_a \pi_{a,v}}\mu_{a,v}^2. \tag{18}$$

Finally, from (16), using (17) and (18), we have

$$Var(\hat{\mu}_{d_1})$$

$$= \frac{1}{N} \left\{ \frac{1 - \gamma_a + \gamma_a \pi_{a,v}}{\pi_a \pi_{a,v}} \sigma^2 + \frac{\gamma_a}{\pi_a} \mu_{a,a}^2 + \frac{(1 - \gamma_a)}{\pi_a \pi_{a,v}} \mu_{a,v}^2 - (\gamma_a \mu_{a,a} + (1 - \gamma_a) \mu_{a,v})^2 \right\}$$

$$= \frac{1}{N} \left\{ \frac{1 - \gamma_a + \gamma_a \pi_{a,v}}{\pi_a \pi_{a,v}} \sigma^2 + \frac{\gamma_a (1 - \gamma_a \pi_a)}{\pi_a} \mu_{a,a}^2 + \frac{(1 - \gamma_a)(1 - (1 - \gamma_a) \pi_a \pi_{a,v})}{\pi_a \pi_{a,v}} \mu_{a,v}^2 \right.$$

$$\left. -2 \gamma_a (1 - \gamma_a) \mu_{a,a} \mu_{a,v} \right\} \tag{19}$$

## A.3 Non-Inferiority Test: Sample Size for Comparing Distinct-Path AIs

Here, we calculate the sample size formula to corresponding to a non-inferiority test to decide whether the AI $d_1$ is non-inferior in efficacy to AI $d_3$. We can write the power of the test as (Friedman et al., 2015)

$$P_{H_1}[\quad reject \quad H_0] = P_{\mu_{d_3} - \mu_{d_1} = \delta} \left[ \frac{\overline{Y}_{d_3} - \overline{Y}_{d_1} - \theta}{\sqrt{v_{d_3 d_1}^{DP}}} < -z_\alpha \right]$$

$$= P_{\mu_{d_3} - \mu_{d_1} = \delta} \left[ \frac{\overline{Y}_{d_3} - \overline{Y}_{d_1} - \delta}{\sqrt{v_{d_3 d_1}^{DP}}} - \frac{\theta - \delta}{\sqrt{v_{d_1 d_3}^{DP}}} < -z_\alpha \right]$$

$$= P \left[ Z < -z_\alpha + \frac{\theta - \delta}{\sqrt{v_{d_3 d_1}^{DP}}} \right], \quad since, \quad Z = \frac{\overline{Y}_{d_3} - \overline{Y}_{d_1} - \delta}{\sqrt{v_{d_3 d_1}^{DP}}} \to N(0,1)$$

$$under \quad H_1$$

$$= \Phi \left( -z_\alpha + \frac{\theta - \delta}{\sqrt{v_{d_3 d_1}^{DP}}} \right),$$

$$or, \quad 1 - \beta = 1 - \Phi \left( z_\alpha - \frac{\theta - \delta}{\sqrt{v_{d_3 d_1}^{DP}}} \right),$$

$$\Phi \left( z_\alpha - \frac{\theta - \delta}{\sqrt{v_{d_3 d_1}^{DP}}} \right) = \beta = \Phi(z_{1 - \beta})$$

$$z_\alpha - \frac{\theta - \delta}{\sqrt{v_{d_3 d_1}^{DP}}} = z_{1-\beta} = -z_\beta$$

$$z_\alpha + z_\beta = \frac{\theta - \delta}{\sqrt{v_{d_3 d_1}^{DP}}} = \frac{\theta - \delta}{\sqrt{(\sigma_{d_3}^2 + \sigma_{d_1}^2)/N}}$$

$$(z_\alpha + z_\beta)^2 = \frac{N(\theta - \delta)^2}{(\sigma_{d_3}^2 + \sigma_{d_1}^2)}$$

$$N = (z_\alpha + z_\beta)^2 \frac{(\sigma_{d_3}^2 + \sigma_{d_1}^2)}{(\theta - \delta)^2}.$$

Note that $\theta = \delta$ means $H_0$ is true, then $N = \infty$. In all other cases, the above formula will give reasonable sample sizes. If $0 < \delta < \theta$, $N$ will be high; if $\delta < 0$, $N$ will be relatively small.

### A.4 Non-Inferiority Test: Variance Formula for Comparing Shared-Path AIs

The variance of $\overline{Y}_{d_3} - \overline{Y}_{d_4}$ is given by

$$V(\hat{\mu}_{d_3} - \hat{\mu}_{d_4}) = V(\hat{\mu}_{d_3}) + V(\hat{\mu}_{d_4}) - 2Cov(\hat{\mu}_{d_3}, \hat{\mu}_{d_4}). \tag{20}$$

Now,

$$Cov(\hat{\mu}_{d_3}, \hat{\mu}_{d_4}) = Cov(\frac{1}{N}\sum_{i=1}^N W_i^{d_3} Y_i, \frac{1}{N}\sum_{i=1}^N W_i^{d_4} Y_i)$$

$$= Cov(\frac{1}{N}\sum_{i=1}^N \frac{I\{T_{1i}=ac, T_{2i}=ac^R i v^{1-R_i}\}}{\pi_{ac} \times \pi_{ac,v}^{1-R_i}} Y_i, \frac{1}{N}\sum_{i=1}^N \frac{I\{T_{1i}=ac, T_{2i}=ac^R i m^{1-R_i}\}}{\pi_{ac} \times (1-\pi_{ac,v})^{1-R_i}} Y_i)$$

$$= \frac{1}{N^2}[E(\sum_{i=1}^N \frac{I\{T_{1i}=ac, T_{2i}=ac^R i v^{1-R_i}\}}{\pi_{ac} \times \pi_{ac,v}^{1-R_i}} Y_i \times \sum_{i=1}^N \frac{I\{T_{1i}=ac, T_{2i}=ac^R i m^{1-R_i}\}}{\pi_{ac} \times (1-\pi_{ac,v})^{1-R_i}} Y_i)]$$

$$- E(\frac{1}{N}\sum_{i=1}^N \frac{I\{T_{1i}=ac, T_{2i}=ac^R i v^{1-R_i}\}}{\pi_{ac} \times \pi_{ac,v}^{1-R_i}} Y_i) E(\frac{1}{N}\sum_{i=1}^N \frac{I\{T_{1i}=ac, T_{2i}=ac^R i m^{1-R_i}\}}{\pi_{ac} \times (1-\pi_{ac,v})^{1-R_i}} Y_i). \tag{21}$$

We have

$$E(\frac{1}{N}\sum_{i=1}^N \frac{I\{T_{1i}=ac, T_{2i}=ac^R i v^{1-R_i}\}}{\pi_{ac} \times \pi_{ac,v}^{1-R_i}} Y_i) = \gamma_A \mu_{ac,ac} + (1-\gamma_A)\mu_{ac,v}, \tag{22}$$

$$E(\frac{1}{N}\sum_{i=1}^N \frac{I\{T_{1i}=ac, T_{2i}=ac^R i m^{1-R_i}\}}{\pi_{ac} \times (1-\pi_{ac,v})^{1-R_i}} Y_i) = \gamma_A \mu_{ac,ac} + (1-\gamma_A)\mu_{ac,m}. \tag{23}$$

Now,

$$E(\sum_{i=1}^{N} \frac{I\{T_{1i}=ac, T_{2i}=ac^{R_i}v^{1-R_i}\}}{\pi_{ac} \times \pi_{ac,v}^{1-R_i}} Y_i \times \sum_{i=1}^{N} \frac{I\{T_{1i}=ac, T_{2i}=ac^{R_i}m^{1-R_i}\}}{\pi_{ac} \times (1-\pi_{ac,v})^{1-R_i}} Y_i)$$

$$= \sum_{i=1}^{N} \sum_{j=1}^{N} E(\frac{I\{T_{1i}=ac, T_{2i}=ac^{R_i}v^{1-R_i}\}}{\pi_{ac} \times \pi_{ac,v}^{1-R_i}} Y_i \times \frac{I\{T_{1j}=ac, T_{2j}=ac^{R_j}m^{1-R_j}\}}{\pi_{ac} \times (1-\pi_{ac,v})^{1-R_j}} Y_j)$$

$$= \sum_{i=1}^{N} E(\frac{I\{T_{1i}=ac, T_{2i}=ac\}}{\pi_{ac}} Y_i \times \frac{I\{T_{1i}=ac, T_{2i}=ac\}}{\pi_{ac}} Y_i)$$

$$+ \sum_{i \neq j} E(\frac{I\{T_{1i}=ac, T_{2i}=ac^{R_i}v^{1-R_i}\}}{\pi_{ac} \times \pi_{ac,v}^{1-R_i}} Y_i \times \frac{I\{T_{1j}=ac, T_{2j}=ac^{R_j}m^{1-R_j}\}}{\pi_{ac} \times (1-\pi_{ac})^{1-R_j}} Y_j)$$

$$= A_{(i=j)} + A_{(i \neq j)}, \tag{24}$$

where,

$$A_{(i=j)} = \sum_{i=1}^{N} E(\frac{I\{T_{1i}=ac, T_{2i}=ac\}}{\pi_{ac}} Y_i \times \frac{I\{T_{1i}=ac, T_{2i}=ac\}}{\pi_{ac}} Y_i)$$

$$= \frac{1}{\pi_{ac}^2} \sum_{i=1}^{N} E(I\{T_{1i}=ac, T_{2i}=ac\} Y_i^2)$$

$$= \frac{1}{\pi_{ac}^2} \sum_{i=1}^{N} [E(I\{T_{1i}=ac, T_{2i}=ac\} Y_i^2 | T_{1i}=ac, T_{2i}=ac) P(T_{1i}=ac, T_{2i}=$$

$$ac) + 0]$$

$$= \frac{1}{\pi_{ac}^2} \sum_{i=1}^{N} [E(Y_i^2 | T_{1i}=ac, T_{2i}=ac) P(T_{1i}=ac) P(T_{2i}=ac | T_{1i}=ac)]$$

$$= \frac{1}{\pi_{ac}^2} \sum_{i=1}^{N} (\sigma^2 + \mu_{ac,ac}^2) \pi_{ac} \gamma_{ac}$$

$$= \frac{N}{\pi_{ac}} \gamma_{ac} (\sigma^2 + \mu_{ac,ac}^2), \tag{25}$$

and,

$$A_{(i \neq j)}$$

$$= \sum_{i \neq j} E(\frac{I\{T_{1i}=ac, T_{2i}=ac^{R_i}v^{1-R_i}\}}{\pi_{ac} \times \pi_{ac,v}^{1-R_i}} Y_i \cdot \frac{I\{T_{1j}=ac, T_{2j}=ac^{R_j}m^{1-R_j}\}}{\pi_{ac} \times (1-\pi_{ac,v})^{1-R_j}} Y_j)$$

$$= \sum_{i \neq j} [\sum_{T_{2i}, T_{2j}} E(\frac{I\{T_{1i}=ac, T_{2i}=ac^{R_i}v^{1-R_i}\}}{\pi_{ac} \times \pi_{ac,v}^{1-R_i}} \cdot \frac{I\{T_{1j}=ac, T_{2j}=ac^{R_j}m^{1-R_j}\}}{\pi_{ac} \times (1-\pi_{ac,v})^{1-R_j}} Y_i Y_j |$$

$$T_{1i} = ac, T_{2i}, T_{1j} = ac, T_{2j}) \times P(T_{1i} = ac, T_{2i}, T_{1j} = ac, T_{2j}) + 0]$$

$$= \sum_{i \neq j} \left[ E\left( \frac{I\{T_{1i}=ac, T_{2i}=ac\}}{\pi_{ac}} \cdot \frac{I\{T_{1j}=ac, T_{2j}=ac\}}{\pi_{ac}} Y_i Y_j \mid T_{1i} = ac, T_{2i} = ac, T_{1j} = ac, T_{2j} = ac \right) \right.$$

$$\times P(T_{1i} = ac, T_{2i} = ac, T_{1j} = ac, T_{2j} = ac)$$

$$+ E\left( \frac{I\{T_{1i}=ac, T_{2i}=ac\}}{\pi_{ac}} \cdot \frac{I\{T_{1j}=ac, T_{2j}=m\}}{\pi_{ac} \times (1-\pi_{ac,v})} Y_i Y_j \mid T_{1i} = ac, T_{2i} = ac, T_{1j} = ac, T_{2j} = m \right)$$

$$\times P(T_{1i} = ac, T_{2i} = ac, T_{1j} = ac, T_{2j} = m)$$

$$+ E\left( \frac{I\{T_{1i}=ac, T_{2i}=v\}}{\pi_{ac} \times \pi_{ac,v}} \cdot \frac{I\{T_{1j}=ac, T_{2j}=ac\}}{\pi_{ac}} Y_i Y_j \mid T_{1i} = ac, T_{2i} = v, T_{1j} = ac, T_{2j} = ac \right)$$

$$\times P(T_{1i} = ac, T_{2i} = v, T_{1j} = ac, T_{2j} = ac)$$

$$+ E\left( \frac{I\{T_{1i}=ac, T_{2i}=v\}}{\pi_{ac} \times \pi_{ac,v}} \cdot \frac{I\{T_{1j}=ac, T_{2j}=m\}}{\pi_{ac} \times (1-\pi_{ac,v})} Y_i Y_j \mid T_{1i} = ac, T_{2i} = v, T_{1j} = ac, T_{2j} = m \right)$$

$$\left. \times P(T_{1i} = ac, T_{2i} = v, T_{1j} = ac, T_{2j} = m) \right]$$

$$= N(N-1)[\gamma_{ac}^2 \mu_{ac,ac}^2 + \gamma_{ac}(1-\gamma_{ac}) \mu_{ac,ac} \mu_{ac,m} + \gamma_{ac}(1-\gamma_{ac}) \mu_{ac,ac} \mu_{ac,v} + (1-\gamma_{ac})^2 \mu_{ac,v} \mu_{ac,m}]$$

$$= N(N-1) \times (\gamma_{ac}\mu_{ac,ac} + (1-\gamma_{ac})\mu_{ac,v}) \times (\gamma_{ac}\mu_{ac,ac} + (1-\gamma_{ac})\mu_{ac,m}). \quad (26)$$

Using, (21) (22), (23), (24), (25) and (26), we have

$$Cov(\hat{\mu}_{d_3}, \hat{\mu}_{d_4})$$

$$= \frac{1}{N^2} \left[ \frac{N}{\pi_{ac}} \gamma_{ac}(\sigma^2 + \mu_{ac,ac}^2) + N(N-1) \cdot (\gamma_{ac}\mu_{ac,ac} + (1-\gamma_{ac})\mu_{ac,v}) \cdot (\gamma_{ac}\mu_{ac,ac} + (1-\gamma_{ac})\mu_{ac,m}) \right]$$

$$- (\gamma_{ac}\mu_{ac,ac} + (1-\gamma_{ac})\mu_{ac,v}) \cdot (\gamma_{ac}\mu_{ac,ac} + (1-\gamma_{ac})\mu_{ac,m})$$

$$= \frac{1}{N} \left[ \frac{\gamma_{ac}}{\pi_{ac}}(\sigma^2 + \mu_{ac,ac}^2) - (\gamma_{ac}\mu_{ac,ac} + (1-\gamma_{ac})\mu_{ac,v}) \cdot (\gamma_{ac}\mu_{ac,ac} + (1-\gamma_{ac})\mu_{ac,m}) \right]. \quad (27)$$

Hence, using ((20)) and ((27)),

$$V(\hat{\mu}_{d_3} - \hat{\mu}_{d_4}) = V(\hat{\mu}_{d_3}) + V(\hat{\mu}_{d_4})$$

$$- \frac{2}{N}\left[\frac{\gamma_{ac}}{\pi_{ac}}(\sigma^2 + \mu_{ac,ac}^2) - (\gamma_{ac}\mu_{ac,ac} + (1-\gamma_{ac})\mu_{ac,v}) \cdot (\gamma_{ac}\mu_{ac,ac} + (1-\gamma_{ac})\mu_{ac,m})\right]. \quad (28)$$

## A.5   Equivalence Test: Power of Comparing Distinct-Path AIs

Here we provide the formula, which can be used to obtain the sample size in an equivalence test that decides whether the AI $d_1$ is equivalent in efficacy to the AI $d_3$. We can write the power of the test as (Friedman et al., 2015)

$$P_{H_1}\left[\frac{\bar{Y}_{d_3} - \bar{Y}_{d_1} - \theta}{\sqrt{v_{d_3 d_1}^{DP}}} < -z_\alpha \quad and \quad \frac{\bar{Y}_{d_3} - \bar{Y}_{d_1} + \theta}{\sqrt{v_{d_3 d_1}^{DP}}} > z_\alpha\right],$$

where $\mu_{d_3} - \mu_{d_1} \in (-\theta, \theta)$ under $H_1$. When $\mu_{d_3} - \mu_{d_1} = \delta \in (-\theta, \theta)$, then the power

$$P_{\mu_{d_3} - \mu_{d_1} = \delta}\left[\frac{\bar{Y}_{d_3} - \bar{Y}_{d_1} - \theta}{\sqrt{v_{d_3 d_1}^{DP}}} < -z_\alpha \quad and \quad \frac{\bar{Y}_{d_3} - \bar{Y}_{d_1} + \theta}{\sqrt{v_{d_3 d_1}^{DP}}} > z_\alpha\right]$$

$$= P_{\mu_{d_3} - \mu_{d_1} = \delta}\left[\frac{\bar{Y}_{d_3} - \bar{Y}_{d_1} - \delta}{\sqrt{v_{d_3 d_1}^{DP}}} - \frac{\theta - \delta}{\sqrt{v_{d_3 d_1}^{DP}}} < -z_\alpha \quad and \quad \frac{\bar{Y}_{d_3} - \bar{Y}_{d_1} - \delta}{\sqrt{v_{d_3 d_1}^{DP}}} + \frac{\theta + \delta}{\sqrt{v_{d_3 d_1}^{DP}}} >\right.$$

$$\left. -z_\alpha\right]$$

$$= P\left[Z < -z_\alpha + \frac{\theta - \delta}{\sqrt{v_{d_3 d_1}^{DP}}} \quad and \quad Z > z_\alpha - \frac{\theta + \delta}{\sqrt{v_{d_3 d_1}^{DP}}}\right]$$

$$= P\left[z_\alpha - \frac{\theta+\delta}{\sqrt{v_{d_3 d_1}^{DP}}} < Z < -z_\alpha + \frac{\theta-\delta}{\sqrt{v_{d_3 d_1}^{DP}}}\right]$$

$$= \Phi\left(-z_\alpha + \frac{\theta-\delta}{\sqrt{v_{d_3 d_1}^{DP}}}\right) - \Phi\left(z_\alpha - \frac{\theta+\delta}{\sqrt{v_{d_3 d_1}^{DP}}}\right), \tag{29}$$

which is equal to $1-\beta$. In a special case, when $\mu_{d_3} - \mu_{d_1} = \delta = 0$, (29) becomes

$$1-\beta = \Phi\left(-z_\alpha + \frac{\theta}{\sqrt{v_{d_3 d_1}^{DP}}}\right) - \Phi\left(z_\alpha - \frac{\theta}{\sqrt{v_{d_3 d_1}^{DP}}}\right) = 1 - 2\Phi\left(z_\alpha - \frac{\theta}{\sqrt{v_{d_3 d_1}^{DP}}}\right)$$

$$\Rightarrow \Phi\left(z_\alpha - \frac{\theta}{\sqrt{v_{d_3 d_1}^{DP}}}\right) = \frac{\beta}{2} = \Phi\left(z_{1-\frac{\beta}{2}}\right)$$

$$\Rightarrow z_\alpha - \frac{\theta}{\sqrt{v_{d_3 d_1}^{DP}}} = z_{1-\frac{\beta}{2}} = -z_{\frac{\beta}{2}}$$

$$\Rightarrow z_\alpha + z_{\frac{\beta}{2}} = \frac{\theta}{\sqrt{v_{d_3 d_1}^{DP}}} = \frac{\theta}{\sqrt{(\sigma_{d_3}^2 + \sigma_{d_1}^2)/N}}$$

$$\Rightarrow \left(z_\alpha + z_{\frac{\beta}{2}}\right)^2 = \frac{N\theta^2}{(\sigma_{d_3}^2 + \sigma_{d_1}^2)}$$

$$\Rightarrow N = \left(z_\alpha + z_{\frac{\beta}{2}}\right)^2 \frac{(\sigma_{d_3}^2 + \sigma_{d_1}^2)}{\theta^2}.$$

## A.6   Data Generation for a SMART

The data generation process for a SMART is complex. Here we outline an algorithm that generates datasets from a SMART (as depicted in Figure 1) to conduct simulation studies. The algorithm is as follows.

1. Set the values of: $Var(Y) = \sigma^2$, $\gamma_{ac}, \gamma_{ac}, \alpha, \beta$.

2.  Set the mean and variance of a latent variable $L$ for the patient with treatment $a$ at first stage as $\mu_{L_a}, \sigma_L^2$, respectively. The latent variable can be viewed as a proximal outcome, which determines responder/non-responder status after the first stage.

3.  Define the cut-off value $\eta$ as that value of the latent variable $L$ above which a patient is considered a responder. So

$$\eta = \mu_{L_a} + \sigma_L \times \Phi^{-1}(1 - \gamma_a),$$

    where $\Phi$ is the cumulative distribution function of the standard normal distribution.

4.  The cut-off $\eta$ should be fixed for a SMART irrespective of the choice of initial treatment ($a$ or $ac$). So we define

$$\mu_{L_{ac}} = \eta - \sigma_L \times \Phi^{-1}(1 - \gamma_{ac}).$$

5.  Define the truncated normal means of the non-responders (NR) who obtained the first-stage treatment $a$ or $ac$ as

$$\mu_{L_{a.NR}} = \mu_{L_a} - \sigma_L \times \frac{\phi(u^*)}{\Phi(u^*)}, \quad where \quad u^* = \frac{\eta - \mu_{L_a}}{\sigma_L}, \quad and$$

$$\mu_{L_{ac.NR}} = \mu_{L_{ac}} - \sigma_L \times \frac{\phi(w^*)}{\Phi(w^*)}, \quad where \quad w^* = \frac{\eta - \mu_{L_{ac}}}{\sigma_L}, \quad respectively.$$

    Here, $\phi$ is the probability density function of the standard normal distribution.

6.  The AI means corresponding to the AIs $d_1 : (a, a^R v^{1-R})$ , $d_2 : (a, a^R m^{1-R})$ , $d_3 : (ac, ac^R v^{1-R})$ and $d_4 : (ac, ac^R m^{1-R})$ can be defined as (see equation (15))

$$\mu_{d_1} = (\zeta_0 + \zeta_{1a}\mu_{L_a}) \times \gamma_a + (\xi_0 + \xi_{1a}\mu_{L_a} + \xi_{2a,v}\mu_{L_{a.NR}}) \times (1 - \gamma_a),$$

$$\mu_{d_2} = (\zeta_0 + \zeta_{1a}\mu_{L_a}) \times \gamma_a + (\xi_0 + \xi_{1a}\mu_{L_a} + \xi_{2a,m}\mu_{L_{a.NR}}) \times (1 - \gamma_a),$$

$$\mu_{d_3} = (\zeta_0 + \zeta_{1ac}\mu_{L_{ac}}) \times \gamma_{ac} + (\xi_0 + \xi_{1ac}\mu_{L_{ac}} + \xi_{2ac,v}\mu_{L_{ac.NR}}) \times (1 - \gamma_{ac}),$$

$$\mu_{d_4} = (\zeta_0 + \zeta_{1ac}\mu_{L_{ac}}) \times \gamma_{ac} + (\xi_0 + \xi_{1ac}\mu_{L_{ac}} + \xi_{2ac,m}\mu_{L_{ac.NR}}) \times (1 - \gamma_{ac}),$$

respectively. Here, $\{\zeta_0, \zeta_{1a}, \zeta_{1ac}, \xi_0, \xi_{1a}, \xi_{1ac}, \xi_{2a,v}, \xi_{2a,m}, \xi_{2ac,v}, \xi_{2ac,m}\}$ are (fixed) parameters to be supplied by the user. Note that $\mu_{a,a} = \zeta_0 + \zeta_{1a}\mu_{L_a}$ shows the dependency of the mean parameter $\mu_{a,a}$ on the mean of the latent variable ($\mu_{L_a}$) using the parameters $\{\zeta_0, \zeta_{1a}\}$ (see equation (15)); the $\mu_{ac,ac} = \zeta_0 + \zeta_{1ac}\mu_{L_{ac}}$ shows the dependency of the mean parameter $\mu_{ac,ac}$ on the mean of the latent variable ($\mu_{L_{ac}}$) using the parameters $\{\zeta_0, \zeta_{1ac}\}$. The $\mu_{a,v} = \xi_0 + \xi_{1a}\mu_{L_a} + \xi_{2a,v}\mu_{L_{a.NR}}$ shows the dependency of the mean parameter $\mu_{a,v}$ on the mean of the latent variables ($\mu_{L_a}$ $and$ $\mu_{L_{a.NR}}$) using the parameters $\{\xi_0, \xi_{1a}, \xi_{2a,v}\}$. Similarly other parameters $\{\xi_{1ac}, \xi_{2a,m}, \xi_{2ac,v}, \xi_{2ac,m}\}$ can be interpreted.

7. Define the effect size between the two distinct-path AIs $d_1$ and $d_3$ as $\delta_{DP} = \mu_{d_3} - \mu_{d_1}$. Similarly, define the effect size between the two shared-path AIs $d_3$ and $d_4$ as $\delta_{SP} = \mu_{d_3} - \mu_{d_4}$.

8. Fix the value $N$ for power curve, otherwise calculate the required minimum value of $N$.

9. Generate $N_a = ceiling(N/2)$ random draws from $N(\mu_{L_a}, \sigma_L^2)$ and generate $N_{ac} = N - N_a$ random draws from $N(\mu_{L_{ac}}, \sigma_L^2)$.

10. Finally generate $N_{T_1,T_2}$ number of primary outcomes $Y$ from $N(\mu_{T_1,T_2}, \sigma^2)$, where

$$\mu_{T_1,T_2} = \begin{cases} \zeta_0 + \zeta_{1a}\mu_{L_a} & \text{if } T_1 = T_2 = a \\ \xi_0 + \xi_{1a}\mu_{L_a} + \xi_{2a,v}\mu_{L_{a.NR}} & \text{if } T_1 = a, T_2 = v \\ \xi_0 + \xi_{1a}\mu_{L_a} + \xi_{2a,m}\mu_{L_{a.NR}} & \text{if } T_1 = a, T_2 = m \\ \\ \zeta_0 + \zeta_{1ac}\mu_{L_{ac}} & \text{if } T_1 = T_2 = ac \\ \xi_0 + \xi_{1ac}\mu_{L_{ac}} + \xi_{2ac,v}\mu_{L_{ac.NR}} & \text{if } T_1 = ac, T_2 = v \\ \xi_0 + \xi_{1ac}\mu_{L_{ac}} + \xi_{2ac.m}\mu_{L_{ac.NR}} & \text{if } T_1 = ac, T_2 = m \end{cases}$$

and

$$N_{T_1,T_2} = \begin{cases} N_a\gamma_a & \text{if } T_1 = T_2 = a \\ N_a(1-\gamma_a) \times \frac{1}{2} & \text{if } T_1 = a, T_2 \in \{v,m\} \\ \\ N_{ac}\gamma_{ac} & \text{if } T_1 = T_2 = ac \\ N_{ac}(1-\gamma_{ac}) \times \frac{1}{2} & \text{if } T_1 = ac, T_2 \in \{v,m\}. \end{cases}$$

If $N_{T_1,T_2}$ contains a decimal part, then $N_{T_1,T_2} = ceiling(N_{T_1,T_2})$.

Appendix Table 1: Parameter values for data generation corresponding to power curve and simulation studies.

| Parameter | Power Curve | Non-inferiority | | Equivalence | | Control of Type-I error rate | |
|---|---|---|---|---|---|---|---|
| | | Distinct path | Shared path | Distinct path | Shared path | Distinct path | Shared path |
| $\theta$ | 2 | {3.0, 2.5} | {3.0, 2.5} | 2 | 2 | 2 | 2 |
| $\sigma$ | 2 | 3 | 3 | 4 | 4 | 4 | 4 |
| $\alpha$ | 0.05 | 0.05 | 0.05 | 0.05 | 0.05 | 0.05 | 0.05 |
| $\beta$ | 0.20 | 0.20 | 0.20 | 0.20 | 0.20 | 0.20 | 0.20 |
| $\gamma_a$ | 0.3 | 0.30 | 0.50 | {0.50,0.45,0.40,0.35,0.30} | {0.50,0.45,0.40,0.35,0.30} | 0.45 | 0.45 |
| $\gamma_{ac}$ | 0.4 | {0.50,0.45,0.40,0.35,0.30} | {0.50,0.45,0.40,0.35,0.30} | {0.50,0.45,0.40,0.35,0.30} | {0.50,0.45,0.40,0.35,0.30} | 0.45 | 0.45 |
| $\mu_{L_a}$ | 2 | 2 | 2 | 2 | 2 | 2 | 2 |
| $\sigma_L$ | 0.2 | 0.2 | 0.2 | 0.2 | 0.2 | 0.2 | 0.2 |
| $\zeta_0$ | 0.02 | 0.02 | 0.02 | 0.02 | 0.02 | 0.02 | 0.02 |
| $\zeta_{1a}$ | 0.5 | 0.8 | 0.8 | 0.5 | 0.5 | 0.5 | 0.5 |
| $\zeta_{1ac}$ | 0.8 | 0.7 | 0.7 | 0.5 | 0.5 | 0.5 | 0.5 |

| $\xi_0$ | 0.03 | 0.03 | 0.03 | 0.03 | 0.03 | 0.03 | 0.03 |
|---|---|---|---|---|---|---|---|
| $\xi_{1a}$ | 0.25 | 0.5 | 0.5 | 0.25 | 0.25 | 0.25 | 0.25 |
| $\xi_{2a,m}$ | 1.3 | 1.3 | 1.3 | 1 | 1 | -1.01 | -1.01 |
| $\xi_{2a,v}$ | 1.3 | {0.01, -0.20} | 0.3 | 1 | 1 | 1 | 1 |
| $\xi_{1ac}$ | 0.5 | 0.4 | 0.4 | 0.25 | 0.25 | 0.25 | 0.25 |
| $\xi_{2ac,m}$ | 0.8 | 0.8 | {-0.38, -0.53} | 1 | 0.94 | 0.95 | 0.95 |
| $\xi_{2ac,v}$ | {3.4,3.2,3.0,2.8,2.6,2.4,2.2, 2.0,1.8,1.5,1.3,1.0,0.8,0.6, 0.4,0.2,0,-0.2,-0.4,-0.6,-0.8,-2.0,-2.2,-2.4,-2.6} | 0.8 | 0.8 | 0.95 | 0.95 | 1 | -1 |

## A.7 Simulated Data Analysis

Here, we demonstrate how the developed methodologies can be applied in practice. In Appendix Table 2, we have four different data sets, indexed 1 through 4. The first data set is generated to answer the primary research question of whether the new adaptive intervention (investigational AI) is non-inferior to the standard adaptive intervention (control AI) with a non-inferiority margin $\theta = 2.5$. Here, the comparison is for two distinct-path AIs and we have $H_0$: New AI is inferior to standard AI, vs. $H_1$: New AI is non-inferior to standard AI.

Based on the prior data, we estimated the standardized effect size to be 0.23. Therefore, the estimated required sample size (N) from Table 2 is 234. Based on the simulated data set 1, the estimated mean outcomes for new and standard AIs are 0.94 and 2.27 respectively (using equation (2)). The variances of the two AIs can be estimated using equation (7). Using the test statistic for comparison of distinct-path AIs, the estimated p-value for the non-inferiority test is 0.0103. Based on the p-value, we reject the null hypothesis and conclude that the efficacy of the new AI is non-inferior to the standard AI with respect to the non-inferiority margin $\theta = 2.5$.

To overcome the over-dependence on the p-value in the hypothesis testing, we have also reported the estimated upper bound of the Bayes Factor, defined as $\dfrac{1}{-e\,p\log(p)}$, where $p$ denotes the p-value (Benjamin and Berger, 2016; Bayarri et al., 2016).

For the non-inferiority test based on the data set 1, the estimated upper bound of the Bayes Factor is 7.81. In other words, the data set 1 implies an odds in favor of the alternative hypothesis (non-inferiority) relative to the null hypothesis (inferiority) of at most 7.81 to 1 (Benjamin and Berger, 2016).

The analysis results for other simulated data sets can be interpreted as above. Note that, for the equivalence test, we need to look at the p-values and upper bounds of the Bayes Factors for both non-inferiority and non-superiority tests to make a decision about the rejection of the null hypothesis. The R-program for this data analysis is available in Open Science Framework (*https://osf.io/hwae3/*).

Appendix Table 2: Data analysis report from four different simulated data sets

|  | Data Set:1 | Data Set:2 | Data Set:3 | Data Set:4 |
|---|---|---|---|---|
| Non-Inferiority (NI) or Equivalence (EQ) | NI | NI | EQ | EQ |
| Comparison of AIs: Distinct (Di)/Shared (Sh) path | Di | Sh | Di | Sh |
| N | 234 | 502 | 277 | 237 |
| Mean Outcome: Treatment AI | 0.94 | 0.39 | 2.4 | 2.62 |
| Mean Outcome: Standard/Control AI | 2.27 | 2.13 | 2.35 | 2.32 |
| p-value: Non-Inferiority Test | 0.0103 | 0.01819 | 0.00243 | 0.00125 |
| Upper bound of the Bayes Factor: Non-Inferiority Test | 7.81 | 5.05 | 25.15 | 44.03 |
| p-value: Non-superiority Test | - | - | 0.00358 | 0.01262 |
| Upper bound of the Bayes Factor: Non-superiority Test | - | - | 18.24 | 6.67 |
| Decision | Reject the null hypothesis | Reject the null hypothesis | Reject the null hypothesis | Reject the null hypothesis |